\documentstyle[emulateapj,onecolfloat]{article}

\def\etal{{\rm et~al.\ }}
\def\hmpc{\;h^{-1}{\rm Mpc}}

\def\hkpc{h^{-1}{\rm kpc}}
\def\kms{{\rm \;km\;s^{-1}}}

\def\kmsmpc{\kms\;{\rm Mpc}^{-1}}
\def\msun{{\rm M_{\odot}}}
\def\ergvol{{\rm\;erg\;s^{-1}\;cm^{-3}}}
\def\ergs{{\rm\;erg\;s^{-1}}}
\def\ergscm{{\rm\;erg\;s^{-1}\;cm^{-2}}}

\def\ergdeg{{\rm\;erg\;s^{-1}\;cm^{-2} deg^{-2}}}
\def\kevkev{{\rm\;keV\;s^{-1}\;cm^{-2}\;sr^{-1} keV^{-1}}}
\newcommand{\PSbox}[3]{\mbox{\rule{0in}{#3}\includegraphics{#1}\hspace{#2}}}

\begin{document}

\twocolumn[

\title{
Hydrodynamic Simulation of the Cosmological X-ray Background
}

\author{
Rupert A.C. Croft$^{1}$,
Tiziana Di Matteo$^{1,5}$,
Romeel Dav\'{e}$^{2}$,
Lars Hernquist$^{1}$,
Neal Katz$^{3}$, 
Mark A. Fardal$^{3}$,
and David H. Weinberg$^{4}$
}

\begin{abstract}
We use a hydrodynamic simulation of an inflationary Cold Dark Matter model with
a cosmological constant to predict properties of the extragalactic X-ray 
background (XRB). We focus on emission from the intergalactic medium 
(IGM), with
particular attention to diffuse emission from warm-hot gas that lies in 
relatively smooth filamentary structures between galaxies and galaxy clusters. 
We also include X-rays from point sources associated with galaxies in the 
simulation, and we make maps of the angular distribution of the emission.  
Although much of the X-ray luminous gas has a filamentary structure, the 
filaments are not evident in the simulated maps because of projection effects.
In the soft (0.5-2 keV) band, our calculated mean intensity of radiation from 
intergalactic and cluster gas is $2.3\times10^{-12} \ergdeg$, $35 \%$ of the 
total soft band emission. This intensity is compatible at the $\sim 1\sigma$
level with estimates of the unresolved soft background intensity from deep
 ROSAT
and {\it Chandra} measurements. Only 4\% of the hard (2-10 keV) emission is 
associated with intergalactic gas. Relative to AGN flux, the IGM component of 
the XRB peaks at a lower redshift (median $z \sim 0.45$) and spans a narrower 
redshift range, so its clustering makes an important contribution 
to the angular
correlation function of the total emission. The clustering on the scales 
accessible to our simulation ($0.1-10$ arcmin) is significant, with 
an amplitude
roughly consistent with an extrapolation of recent ROSAT results to small 
scales. A cross-correlation analysis of the XRB against nearby galaxies taken 
from a simulated redshift survey also yields a strong signal from the IGM.
Our conclusions about the soft background intensity differ from those of some
recent papers, which have argued that the expected emission from gas in galaxy,
group, and cluster halos would exceed the observed background unless much of
the gas is expelled by supernova feedback. 
We obtain reasonable compatibility with
current observations in a simulation that incorporates cooling, star formation,
and only modest feedback.
A clear prediction of our model is that 
the unresolved
portion of the soft XRB will remain mostly unresolved even as observations
reach deeper point-source sensitivity.
\end{abstract}
 
\keywords{Cosmology: observations,
 large scale structure of Universe}
 
]

\footnotetext[1]{Harvard-Smithsonian Center for Astrophysics, 
Cambridge, MA 02138; rcroft,tdimatte,lars@cfa.harvard.edu}
\footnotetext[2]{Princeton University Observatory, Princeton, NJ 08544;
 rad@astro.princeton.edu}
\footnotetext[3]{Department of Physics and Astronomy, 
University of Massachusetts, Amherst, MA, 01003;
nsk@kaka.astro.umass.edu,fardal@weka.astro.umass.edu}
\footnotetext[4]{Department of Astronomy, The Ohio State University,
Columbus, OH 43210; dhw@astronomy.ohio-state.edu}
\footnotetext[5]{{\it Chandra} Fellow}

\section{Introduction}
It is now well established that the Cosmic X-ray Background (XRB)
results almost entirely
from the integrated X-ray emission of many discrete sources.
Deep ROSAT surveys have shown that a large fraction of the soft ($0.5-2$ 
keV)  XRB is produced by active galactic nuclei (AGN), i.e. quasars and
Seyfert 1 galaxies (e.g.,~Hasinger et al. 1998; Schmidt et al.~1998).
Until recently,
the nature of the sources producing the energetically dominant, hard
($2-10$ keV) XRB was largely unknown. 
{\it Chandra} has now imaged the hard
(2-10 keV) X-ray sky at high resolution and resolved essentially
all of the XRB in this band (down to a flux of $4 \times 10^{-15}
\ergscm$; Mushotzky \etal 2000; see also Brandt et al. 2000). The
optical follow-up campaigns have shown that the newly discovered XRB 
sources include (1) optically `faint' galaxies, which may be either 
quasars at high redshifts or obscured AGN, and (2) apparently `normal'
galaxies.

However, a non-negligible fraction of the sky brightness, particularly at 
soft energies, may still be contributed by diffuse emission 
(mainly Bremsstrahlung) from intergalactic gas. Cosmological 
structure formation theories have been used to estimate the 
radiation emitted by gas in galaxy clusters
 (e.g., Blanchard \etal 1992). In the last few years,
attention has also focused (Cen \etal 1995; 
Ostriker \& Cen 1996; Dav\'{e} \etal 2000, hereafter D00)
 on the large fraction of 
baryonic matter at low redshift predicted by these theories
to  be in the form of a ``warm-hot'' ($10^{5}-10^{7}$K) intergalactic
medium (WHIM). 
As Cen \& Ostriker (1999, hereafter CO99) have emphasized,
detection of the faint, soft-band X-ray emission from this
currently unobserved gas would help reconcile measurements 
of the baryon density in the local universe (e.g., Fukugita,
Hogan \& Peebles 1998) with the predictions 
of cosmological nucleosynthesis (Burles \& Tytler 1998). 
In this paper, we use a cosmological hydrodynamic simulation to 
predict the intensity and angular distribution of X-ray emission from 
intergalactic gas. Our approach includes this relatively low
density filamentary component, as well as clusters and groups.
We also include a simple prescription for emission from AGN in 
the simulation.

Hydrodynamic simulations are routinely used to examine the
expected X-ray emission from galaxy clusters in a cosmological context
(see, e.g., Frenk \etal 1999 and references therein). The lower density gas 
in structures between clusters in such simulations 
is morphologically more complicated, being distributed in the usual 
filaments and sheets. One can ask what observational
signatures of X-ray emission are expected from this gas.
For example, are the filaments
visible in projection  on the sky? Experience with filamentary structure
in the galaxy distribution, which only stands out clearly in
3-dimensional redshift maps (e.g., de Lapparent, Geller \& Huchra 1986),
suggests that they will be difficult to detect.
Nevertheless, some evidence has been found for diffuse filamentary 
emission in ROSAT deep
pointings (Scharf \etal 1999). Images of individual
filaments made from simulations by Pierre, Bryan \&
Gastaud (2000)
suggest that the {\it XMM-Newton} telescope should be able to detect
them, at least when they are considered in isolation.
An alternative approach, which we undertake here, is to
make simulated sky maps using the cosmological models and measure statistical
properties of the X-ray flux, which can eventually be compared to 
observations.

Another important question is the integrated intensity of the
background.  Several recent papers have claimed that the expected
emission from gas in groups of galaxies or galaxy halos
exceeds the observed diffuse
soft X-ray background by at least an order of magnitude (Pen 1999; Wu,
Fabian, \& Nulsen 1999).  They suggest that much of 
the gas in these groups and galaxies halos must be expelled or inflated
by supernova
feedback, decreasing its density and hence its X-ray luminosity.
The amount of supernova feedback required is quite extreme, approaching
1 keV per baryon.
However, these these calculations are not based on
self-consistent simulations including cooling and galaxy formation.
CO99 have carried out a
large Eulerian hydrodynamic simulations which does include these
processes, incorporating supernova feedback much milder than
that advocated by Pen, and Wu, Fabian \& Nulsen. CO99 find
that the intensity of the 0.7 keV background predicted by
their simulation is in fact consistent with observations. 
In this paper, we use a different numerical technique to CO99, but also 
only include very modest supernova feedback
It is important to see whether our simulation gives similar results
for the mean XRB intensity.

The layout of the paper is as follows. In \S2, we present the
hydrodynamic simulation and cosmological parameters used.
We examine
the diffuse X-ray emission from the IGM in \S3, and study the relationship
between X-ray emissivity and gas density.
In \S4, we describe our procedure
for associating X-ray luminosities with some galaxies in the simulation.
Simulated sky maps are generated as described in \S5, and their clustering
is analyzed in \S6. We summarize and discuss our results in \S7.
Finally, in  Appendix A, we explain how and why we account for the multiphase
nature of the gas when calculating its X-ray emissivity,
and carry out some tests of our procedure.

\section{The Simulation}

D00 have recently studied the distribution of
gas temperatures and densities in several simulations of Cold Dark Matter
models. One of these is the simulation that we use here, their ``D1''.
We refer the reader to their paper
for more details on this simulation, 
as well as a comparison of the effects of resolution,
box size, and numerical technique
on the fraction of gas in different phases.

The simulation is of a cosmological constant dominated Cold Dark Matter
model, run using the code Parallel TreeSPH (Dav\'{e}, Dubinski \& Hernquist
 1997). The parameters used are $\Omega_{M}=0.4$, $\Omega_{\Lambda}=0.6$,
a Hubble constant $H_{0}= 65\;\kmsmpc$, an amplitude
of mass fluctuations $\sigma_{8}=0.8$, an initial
spectral index $n=0.95$, and a baryon fraction
$\Omega_{b}= 0.02h^{-2}$ (Burles \& Tytler 1998). The simulation volume
is a cube of side-length $50 \hmpc$ and initially contained
 $144^{3}$ dark matter
particles and as many gas particles, which
results in a baryonic mass resolution of 
$8.5 \times 10^{8} \msun$
per particle. The spatial resolution of the simulation
is 7 comoving $\hkpc$ (equivalent Plummer softening), and it was evolved from
$z=49\rightarrow 0$.
A prescription for star formation converts gas particles in convergent
flows in cooling, dense
regions into star particles and deposits feedback energy from 
supernovae into the surrounding gas (see Katz, Weinberg \&
Hernquist 1996 for details).
Because these regions are so dense, the energy rapidly radiates
away and has only a limited effect on the large-scale gas morphology.
Feedback from stars
in the simulation is therefore relatively unimportant, something that 
should be borne in mind when considering the X-ray emission
from clusters and other dense regions. We discuss this point further below.

\begin{figure}[t]
\centering
\PSbox{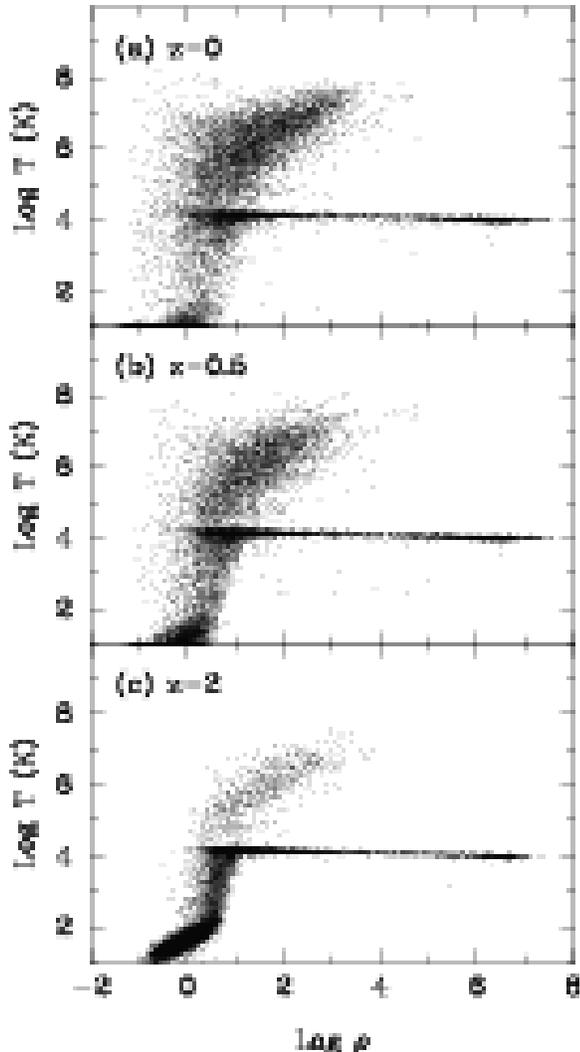 angle=-90 voffset=440 hoffset=-195 vscale=80 hscale=80}
{3.5in}{5.5in} 
\caption
{
The temperature and density of gas particles in the simulation 
at 3 different redshifts. We plot a randomly chosen 10000 particles
in each panel.
\label{rhot}
}
\end{figure}

In Figure \ref{rhot}, we plot the temperatures and densities of the
gas particles in the simulation at three different redshifts.
D00 have found, in their comprehensive study, that  
at the present day, $30-40 \%$ of baryons are predicted 
to be in a WHIM, with 
a relatively low median overdensity of $\sim 20 $ times the cosmic mean. 
In our calculation of the X-ray emission, all the gas above $T \sim 10^{5}\;$K
will make a contribution, with the IGM emission being
dominated by the hottest cluster gas.

We can see from Figure \ref{rhot} that the fraction of hot gas increases
with decreasing $z$, as more gas becomes shocked and heated by falling into high
density regions (the fraction above $10^5$ K climbs from $5\%$ at $z=2$
to $28\%$ at $z=0.5$ and $41 \%$ at $z=0$).The three 
phases of gas identified by, e.g., Dav\'{e} \etal (1999a) are clear
from the Figure: a cold diffuse IGM, a shocked plume of hot gas, 
and a cold, condensed, collapsed fraction. 
We note that in this simulation  no ionizing background
radiation has been included, so that the lowest density gas is all at a very
low temperature.
The densities in the plot have been subjected to a correction
to make allowances for the co-existence of two gas phases ---
hot/diffuse and cold/dense --- in galaxy groups and clusters.
The SPH algorithm tends to blur the boundary between these phases
when they are in close proximity and the mass resolution is limited,
and it is necessary to separate them explicitly 
to properly calculate the X-ray emission.
We describe this correction in detail in Appendix A, and also
briefly in \S3 below.

\section{Diffuse X-ray emission}

The IGM is a hot, thin plasma that 
emits X-rays via a number of physical mechanisms,
including Bremsstrahlung, collisional excitation of spectral lines,
and recombination. Metal line emission is  most important
for low density, cooler gas, while thermal Bremsstrahlung  dominates the
emission from the intracluster medium.
We use a Raymond-Smith  (1977) code to calculate the X-ray emission from the
IGM. Given the temperature, metallicity, and electron density of
gas, the code calculates the volume emissivity.
In the maps that we make, much of the radiation will come from
relatively high redshift, so that redshifting into 
the observed band is important. 

As the metallicity of the gas is not tracked self-consistently in the 
simulation, we assign a metallicity that is a simple function
of  density. For gas at the mean density, the metallicity 
we assign is 0.005 of the solar value, with the metallicity being 
proportional to $\sqrt{\rho}$
 in other regions and being limited to a maximum of 0.3
Z$_{\odot}$. In this way, we roughly mimic the simulation result of
Cen \& Ostriker (1999), which in turn 
 approximately reproduces the fraction of metals
found in the low density IGM at high redshift, and the metallicity of
groups and clusters in the Local Universe. 
Another study of
metal enrichment, involving several different
physical models, 
has been carried out by Aguirre \etal (2000ab), starting from the same
hydrodynamic simulation that we use here. 
We leave the detailed study of the effect of metal enrichment
on X-ray emission (and absorption, see, e.g., Hellsten,
Gnedin \& Miralda-Escud\'{e} 1998 and Perna \& Loeb 1998)
 to future work.  
In the present paper, we limit our study of varying metallicity
to testing only the effect of removing all emission from metals.

Much of the radiation from the IGM will be in the form of relatively soft
X-rays, corresponding to thermal emission from
filaments, groups, and clusters,
which have temperatures  that range
from a fraction of 1 keV to a few keV. We will make maps of emission
in two separate  bands, $0.5-2$ keV (soft) and $2-10$ keV (hard),
spectral intervals motivated by those  commonly used in the analysis of
observational data.
The hard band will therefore contain emission
representative of only a small portion of the IGM, the hottest clusters.
Below 0.5 keV, extragalactic emission becomes impossible to separate
cleanly from  
galactic coronal emission, and galactic absorption 
becomes even more important. Because of this, we do not attempt 
to make maps or study clustering of the very soft XRB. We will, however,
calculate the mean intensity in a band centered on 0.25 keV, due to the
WHIM, which can be compared to X-ray shadowing measurements 
(e.g., Wang \& Ye 1996; Cui \etal 1996).

\subsection{The X-ray luminosity of simulation gas particles}

The emissivity of gas is given in $\rm erg\;s^{-1}$ per unit volume
by the Raymond-Smith code. 
The simulation we have is Lagrangian, so that the 
gas is discretized into particles rather than
cells of unit volume. 
To calculate the X-ray luminosity of each particle
we must therefore calculate the volume of gas associated with it. 
First, the density for each particle is calculated using the symmetrized
SPH spline kernel (see Hernquist \& Katz 1989), with 32 neighbors.
The volume associated with the particle is  
\begin{equation}
V=\frac{0.76 
M_{g}}{m_{H}n_{H}},
\label{volpar}
\end{equation}
where $M_{g}$ is the particle gas mass, $m_{H}$ the
mass of a hydrogen atom, and $n_{H}$ the number density of
hydrogen atoms (which are fully ionized in nearly all
the hot gas we are considering).
We take the fraction by mass of helium in primordial gas
to be 0.24, which accounts for the factor of 0.76 for hydrogen
(we use the same formula for gas of non-primordial
composition, ignoring the small contribution to the mass made by metals).  

A small number of the
hot gas particles in clusters and groups will have many neighbors 
which are cold, condensed galactic gas. The intracluster
gas is, however, in a separate phase, so that for the SPH density estimate to
have physical meaning, we choose not to include the contribution
from the cold, dense gas. Without separating out the cold galactic gas,
the density of these hot particles will be overestimated. 
With higher mass resolution, this would not be such a problem, as
the different phases would be better resolved. In the present 
simulation, however, this correction to the densities of the hot particles
makes a large difference to the overall X-ray emission.
In equation~(\ref{volpar}), therefore, we
estimate $n_{H}$ for hot ($T>10^5\;$K) particles
after removing particles that have $T < 10^{4.5}$ K
and densities greater than $1000$ times the mean. We use this density
estimate in Figure \ref{rhot} and in the rest of the results in the
main text of this paper.

In Appendix A, we describe the motivation for this
corrected density estimate in greater detail, and we perform a number
of tests to show that it gives correct results.
For previous approaches to dealing with this problem, in the context
of galaxy formation, the reader is referred to Pearce \etal (1999)
and Ritchie \& Thomas (2000).
Also, Pearce \etal (2000) have calculated 
the X-ray luminosities of galaxy clusters
from simulations run using a similar multiphase density estimate.

\subsection{Emissivity and density}

\begin{figure}[t]
\centering
\PSbox{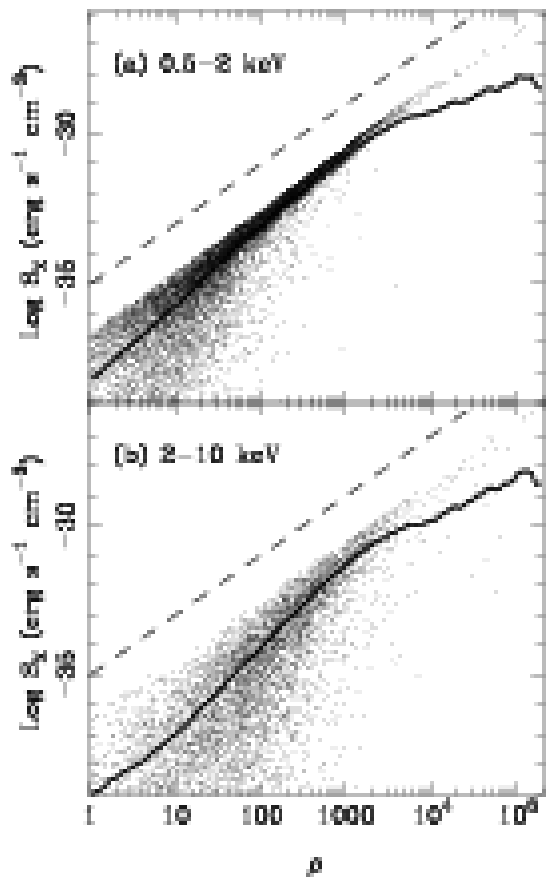 angle=-90 voffset=395 hoffset=-185 vscale=75 hscale=75}
{3.5in}{4.5in} 
\caption
{
Emissivity vs. gas density (in units of the cosmic mean) at $z=0$
in two bands (a) soft, (b) hard. The solid line is the
average emissivity at a given density. The dashed line represents
 $S_{X}\propto\rho^{2}$, with arbitrary  normalization.
A randomly chosen $1\%$ of the particles is plotted.
\label{rhosx}
}
\end{figure}

In Figure \ref{rhosx}, we plot the volume emissivity of particles against
their density in units of the mean. We find that some X-ray emission
comes from gas at low densities, at or around the cosmic mean, which is
probably recently shocked gas. Most of the X-ray emitting particles 
have densities in the range $\rho=100-1000$. The
Bremsstrahlung  component of the emissivity is
proportional to $\rho^{2}T^{0.5}$. The temperature is roughly 
proportional to the density in the hot diffuse gas regime (although
with quite large scatter, see Figure \ref{rhot} and figure 6 of
D00). This results in a mean emissivity at a given density
that is approximately a power-law with a slope of 2.5. 
At very high densities, the solid
line, which represents the mean emissivity,
bends downwards due to the large fraction of particles that have cooled. 
The scatter about the line increases at lower densities, with the
slope of the upper envelope being a roughly 
constant $S_{X}\propto \rho^{2}$. This larger scatter at lower
densities may be due to the widely differing histories of 
gas elements at low density, where shock heating has occurred to different
extents. At higher densities, gas particles tend to relax
more uniformly to the virial temperature.

\begin{figure}[t]
\centering
\PSbox{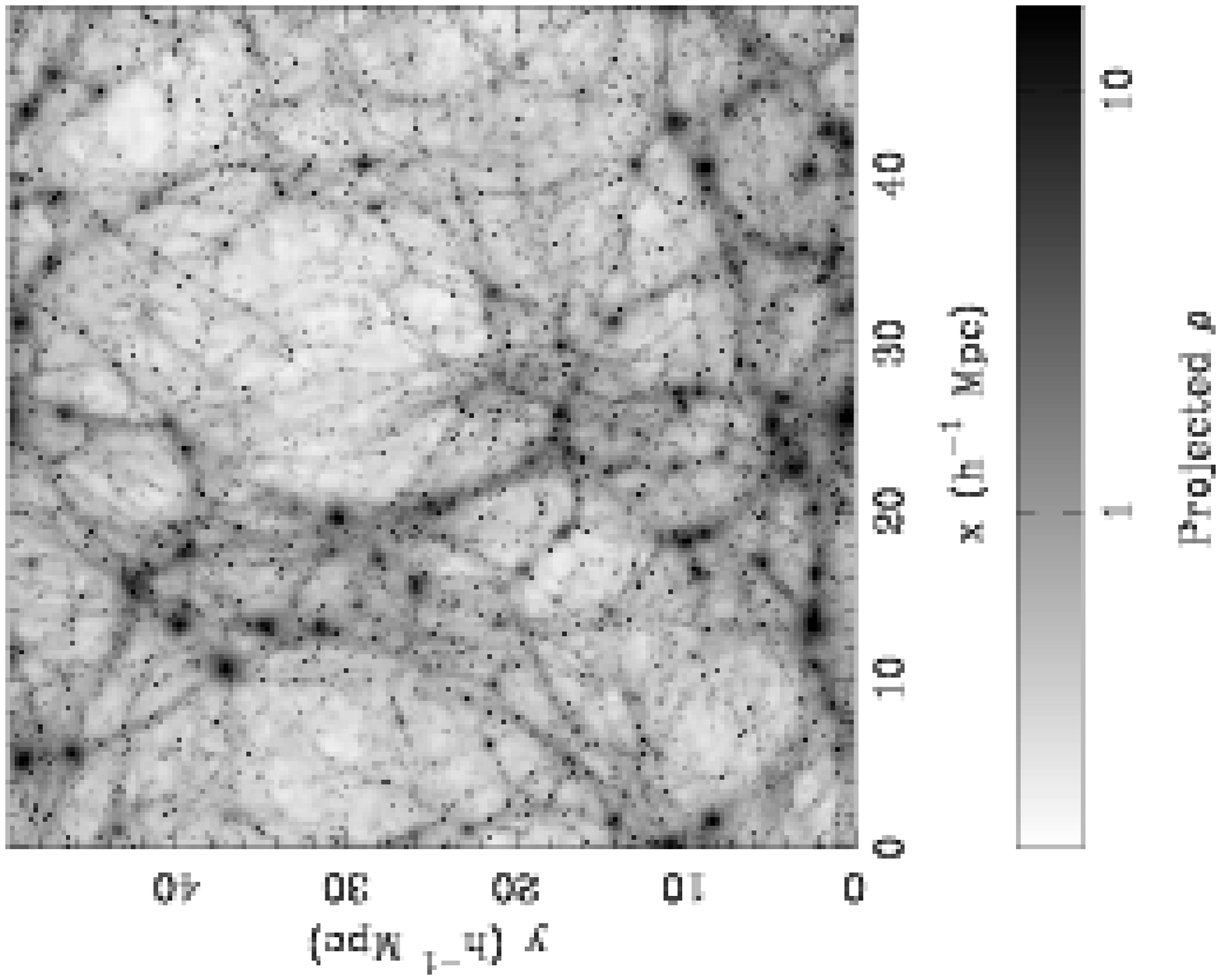 angle=-90 voffset=325 hoffset=-100 vscale=55 hscale=55}
{3.5in}{4.1in} 
\PSbox{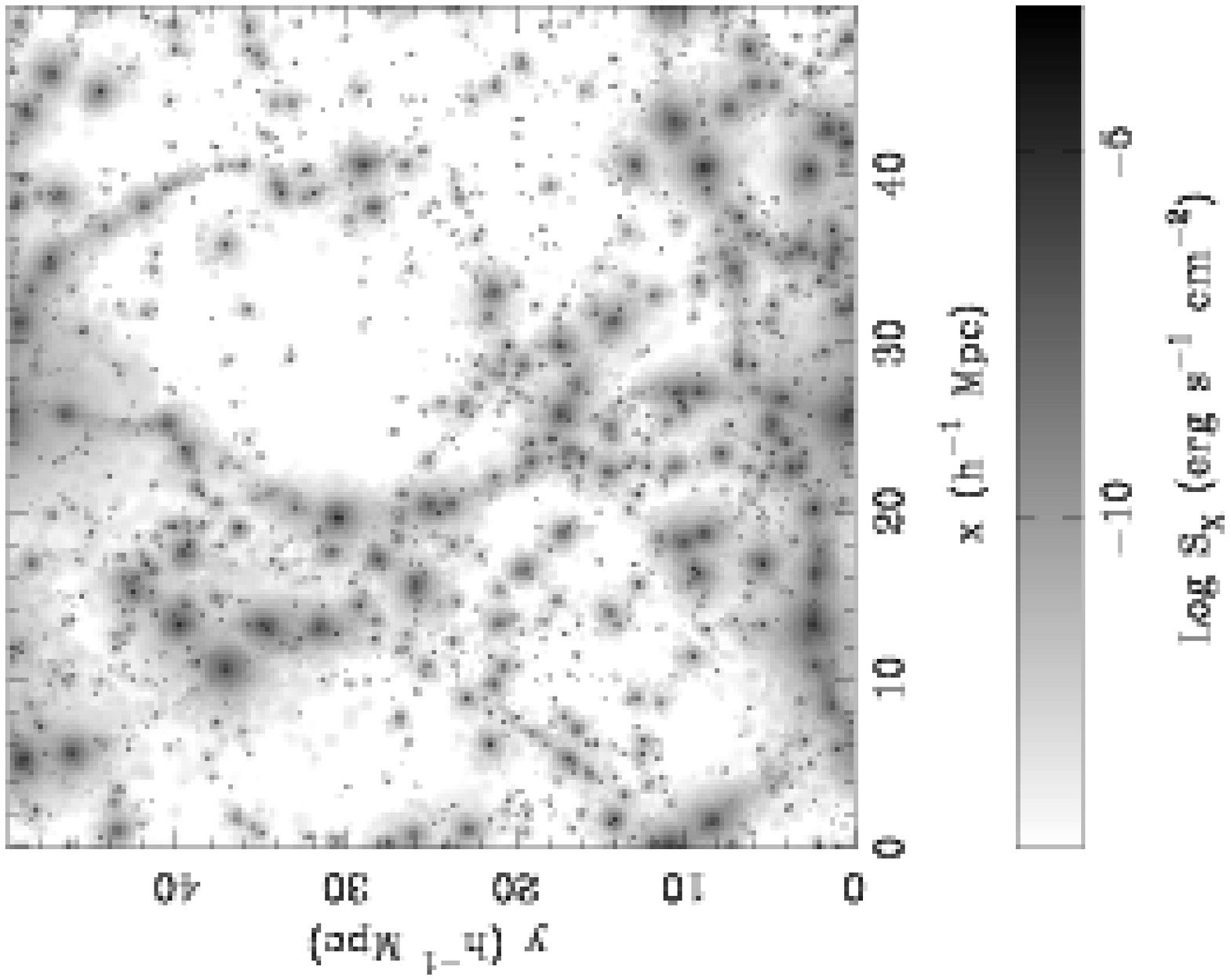 angle=-90 voffset=325 hoffset=-100 vscale=55 hscale=55}
{3.5in}{4.1in} 
\caption
{
Projection of density (top) and X-ray emission in the soft band (bottom) at
$z=0$.
\label{slice}
}
\end{figure}

We examine the spatial distribution of gas density and X-ray emission at $z=0$
in Figure \ref{slice}. To treat both in a similar way, we
assign the gas mass and the X-ray emission to a two-dimensional
grid using the projected SPH kernel. The X-ray emissivity per unit
area in grid cell $jk$ is therefore 
(see also Springel, White, \& Hernquist 2000)
\begin{equation}
\frac{1}{A}\sum_{i=1}^{N} {}^{X}S_{i} W_{i,j,k}
\label{spha}
\end{equation}
where $N$ is the number of particles, and $^{X}S_{i}$ is
the X-ray flux emitted by particle $i$
(see also Springel, White, \& Hernquist 2000).
$W_{i,j,k}$ is the integral of the projection of the
SPH smoothing kernel in the $x-y$ plane over grid cell $j,k$.
$A$ is the area of a grid cell. For the mass density assignment,
we replace $^{X}S_{i}$ with the gas mass of particle $i$.
By using the SPH kernel in this way, we make use of the fact that we have 
higher spatial resolution in higher density regions. We use the same procedure
in the sky maps that we present in \S5.

In the case of Figure \ref{slice}, because we have projected the entire
simulation
volume, the absolute units on the greyscale are not very useful. In relative
terms, though, one can see that the dynamic range of the X-ray scale is much 
larger than for the mass, as one might expect, given that
$S_{X} \propto \sim \rho^{2.5}$. X-ray filaments are therefore likely to
require very faint imaging to be picked out (see Pierre et al.\ 2000
for simulated {\it XMM-Newton}
imaging of individual filaments). Comparing the two maps,
it is also evident that much of the fine structure and many of the small 
filaments seen in the mass distribution do not show up in the X-rays.
The X-ray filaments seem to be even more ``beaded'' than the matter
structures, being made up of small galaxy groups lying
close together. There are also many smaller clumps of gas emitting
X-rays that lie in relative isolation, in regions of lower density.
We can see in the density plot that many of these are associated with 
individual galaxies 
and the emission is from gas in their halos. 
The X-ray luminosity function of the galaxy groups will be studied in
a future paper.

\section{Active galactic nuclei}

As mentioned in \S1, AGN are the major contributors to the XRB.
The hydrodynamic simulation we are using was run primarily to
study galaxy formation (Dav\'{e} \etal 1999b; Weinberg
\etal 1999), and it therefore includes galaxies that we can use
to model the AGN contribution.

 The recent dynamical evidence for supermassive black holes in 
many nearby galaxies, and the correlations between the mass of the
central black hole and the mass (Magorrian et al. 1998) or 
velocity dispersion (Ferrarese \& Merritt 2000; Gebhardt \etal 2000)
of its host bulge,
provide strong support for the idea that most galaxies should contain
central supermassive black holes. We therefore associate a central
X-ray point source with each galaxy in the simulation and include the
radiation emitted by these
AGN in our  X-ray maps.

 Since AGN are believed to
be powered by accretion onto black holes, a process far below the
resolution of the simulation, we will necessarily have to treat the
X-ray emission from black holes in the simulated galaxies in a
simplified fashion. In the future, it may become possible to track 
some aspects of the
merging and growth of supermassive black holes 
self-consistently in hydrodynamic simulations
(see, e.g., Merritt, Cruz, \& Milosavljevic 2000 for preliminary
work along these lines).

 For now, the primary role for our simulated AGN
population is to be placed in our skymaps along with the more
accurately computed diffuse emission, so that we can approximately
gauge their effect on the overall clustering and flux level of the
XRB. Even so, there is some information that can be gained about
the likely nature and distribution of AGN in a CDM universe from our
use of a hydrodynamic simulation.

Emission from AGN has previously been included in the hydrodynamic
simulations of Cen \& Ostriker (e.g.,  Cen \& Ostriker 1993).
Our study also represents an attempt to
incorporate a simple scheme for AGN and their X-ray emission
into the framework of
hydrodynamic simulations.

 In recent work, Kauffmann \& Haehnelt (2000,
hereafter KH), Cattaneo (1999), and Wilman, Fabian \& Nulsen (2000,
hereafter WFN) have used semi-analytic galaxy formation models and
associated prescriptions for quasar fueling to study the growth of
supermassive black holes. Models by KH and Cattaneo assume that black
holes are fed by cold gas during major galaxy mergers. This allows
them to reproduce the B-band luminosity function of quasars, 
but they do not use the XRB as a constraint.  WFN instead assume
that quasars are fueled by Bondi accretion from the dense, hot gas
phase in galaxies, and they attempt to reproduce the hard XRB and X-ray
source counts.

 Our approach is complementary to (albeit less ambitious than)
the work in these three papers. 
We will assume that a fraction of the cold gas in each galaxy
is accreted by the central
black hole and calculate its X-ray emission using current
constraints on AGN X-ray spectra.

\subsection{Galaxy selection}
 
We use the SKID groupfinder (see, e.g., Governato \etal 1997) to find
distinct clumps of gas and star particles. The minimum number of
particles in a clump (including both cold gas and stars) before it is
considered a galaxy is set to be 4, with a corresponding mass of
$3.4\times10^{9} \msun$.  At $z=0$, there are 5500 of these galaxies
in the simulation volume, although at $z=2$, around the peak of the
AGN distribution, there are only $210$. We use a small mass limit 
to have as large a number of galaxies as possible. Below
$\sim60$ particles, the list of objects above a given mass will be
incomplete. As we will in any case be sparsely sampling the galaxy distribution
to find AGN and assigning their properties with parameterized prescriptions,
this incompleteness is relatively unimportant for our purpose.
Although the limit of 4 particles might seem low, 
the overdensity threshold for star formation is $\rho>1000$, 
and cold gas must also satisfy this threshold to be counted in a galaxy,
so the existence of a 4-particle SKID group implies the existence
of a larger dense system surrounding it.
 
For each galaxy the simulation provides us with
$M_{gas}$, the amount of cold gas in the galaxy, 
and $M_{star}$, the mass in stars. 
The ratio $M_{gas}/M_{star}$ is generally higher at higher redshift.
We also have the spatial positions of the galaxies,
and their halo circular velocities, $V_{c}$.

\subsection{Emission from AGN}
 
\begin{figure}[t]
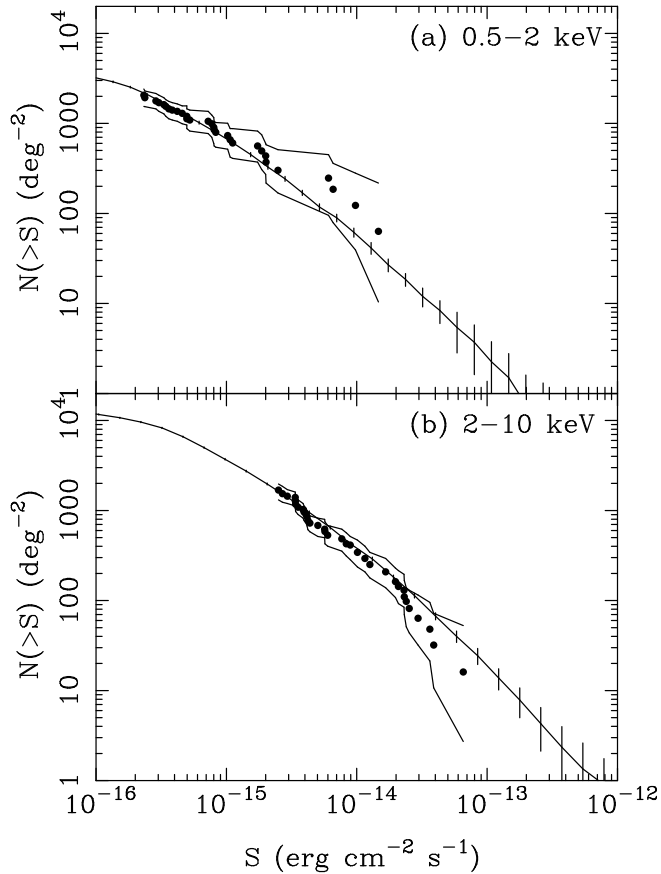

\centering
\PSbox{lognlogs.ps angle=-90 voffset=360 hoffset=-120 vscale=65 hscale=65}
{3.5in}{4.7in} 
\caption
{
Number counts ($\log N - \log S$)
for the simulated AGN (smooth solid line).
The variance in results
from separate survey areas of 1 degree is shown by the error
bars.
The {\it Chandra} results of Mushotzky 
\etal (2000) are shown as points, with their $1 \sigma$
statistical errors denoted by jagged lines.
\label{lognlogs}
}
\end{figure}
 
We use this galaxy information together with an approximate physical model to
simulate the point-source X-ray emission. As mentioned above, we are
forced to include somewhat {\it ad hoc} relationships between variables and
free parameters. 
 
By analogy with the prescriptions adopted in semi-analytical models,
e.g. KH or Cattaneo, Haehnelt, \& Rees (1999, hereafter CHR), we assume 
that the ratio of the accreted mass to total cold gas
scales with circular velocity:
\begin{equation}
M_{acc} = f M_{gas}\left(\frac{V_c}{v_{\rm max}}\right)^4 ~,
\label{macc} 
\end{equation}
where $V_c$ is the halo circular velocity, and $v_{\rm max} = 200\kms$,
$f = 10^{-4}$ are free parameters chosen so that
the simulated population approximately reproduces the
observed AGN X-ray luminosity function at $z
\sim 0$ (e.g., Boyle et al.~1998).

Another constraint on these parameters, as well as on
the power-law exponent of $V_{c}$, is
the source count measurement by the {\it Chandra} satellite.
The $\log N-\log S$ results of Mushotzky \etal (2000)
are shown in Figure \ref{lognlogs}. 
The fourth power dependence on $V_{c}$ 
is stronger than that adopted by KH, but we find that such a dependence
is necessary to reproduce the observational results. We
will return to this point below.
The strong
dependence on velocity 
in Equation \ref{macc} may be supported by the recent work of
Ferrarese \& Merritt (2000) and Gebhardt \etal (2000),
who find $M_{BH} \propto \sigma_{v}^{\sim 4}$ ($\sigma_{v}$ is 
bulge velocity dispersion) in nearby galaxies.
In the framework of our model, this would imply that
the mass of the black hole is
built up mostly by accretion ($M_{BH} \propto M_{acc}$).

The major active phase takes a short fraction of a galaxy's
lifetime.  This mass is accreted on a timescale $t_{acc}(z) \sim 1 \times
10^{7} (1+z)^{-2}$ yr, which we can also refer to as the quasar lifetime.
The $z$ dependence in the timescale is necessary to
reproduce the peak of the quasar phase and the fast fall off at $z <2$
(see also KH and CHR).  With this assumption, $\dot{M} = M_{acc}/t_{acc}$, we
take the
probability of a given AGN being active at redshift $z$
 to be $P_{on}
t_{acc}(z)|dt_{H}(z)/dz|^{-1}$, where $t_{H}(z)$
is the Hubble time at redshift $z$. $P_{on}$ is another
free parameter, necessary because we must randomly sample the
galaxy population, as we use the same simulation at different redshifts
rather than following the evolution of individual galaxies. 
We do not specifically treat the effects of obscuration in
AGN. The presence of a population of obscured AGN (at $z > 2$) can be
relevant for producing significant quantities of hard X-rays. The
properties of such sources, however, are required to be quite different
from those of the established local classes of AGN, and their modeling
would imply the introduction of further assumptions for the
prescription of the absorbing material (see, e.g., WFN).
 
We take the luminosity of the nuclear source to be $L_{bol}= \eta
\dot{M} c^2$.  The 2-10 keV X-ray luminosity in AGN is typically equal
to 3\% $L_{bol}$ (Elvis et al.~1994). We assume a standard accretion
disk efficiency of $\eta = 10\% $ as long as the accretion rate
exceeds $\sim 1.3 \alpha^2 \dot{M}_{Edd}$ (where 
$\dot{M}_{Edd} \equiv L_{Edd}/(0.1c^2)$ and we assume $\alpha=0.1$
for the disk viscosity parameter). When the accretion
rate is lower than this, we set $\eta = 10^{-3}$ to allow for an
advection-dominated, low luminosity phase (which turns out to be
important only at $ z < 1$).  We set the black hole masses using the
Magorrian \etal (1998) relationship (valid at these low redshifts)
between bulge mass and black hole mass, assuming a bulge mass
proportional to the total galaxy mass in the simulation.

  The X-ray spectrum of a typical
AGN ($\eta = 0.1$) is a power-law with photon index $\Gamma =1.8$,
and with an exponential cut off at 200 keV (irrelevant for 0.5 -- 10 keV
energies). The X-ray spectrum of the fainter sources ($\eta =10^{-3}$)
is harder, with a photon index $\Gamma =1.4$, as expected from 
Bremsstrahlung emission from hot, radiatively inefficient accretion
flows (and similar to that of the hard X-ray background).

Given the above model, we
calculate $\log{N}-\log{S}$ for the point-source
population (Figure \ref{lognlogs}).
The free parameters in the model are
those in equation~(\ref{macc}), which govern the luminosity of
each AGN ($f$, $v_{max}$, and the $V_{c}$ exponent), and are
responsible for the shape and left-right movement of the  $\log{N}-\log{S}$
curve. The other parameter, $P_{on}$,
multiplies the probability of a quasar being on, and moves
the $\log{N}-\log{S}$ curve up and down.
It is degenerate in this respect with $t_{acc}$, the quasar lifetime,
which we choose not to vary.
As mentioned above, we have adjusted the 
free parameters so that our results approximately reproduce the
$\log{N}-\log{S}$ measured by Mushotzky \etal (2000) (also
shown in Figure \ref{lognlogs}), and the $z=0$ QSO X-ray luminosity
function (Boyle \etal 1998, not plotted). 

One point to note is that equation~(\ref{macc}), which governs
the relationship between the black hole accretion rate and
halo circular velocity, is required to be steep to
reproduce the relatively shallow slope of the observed $\log{N}-\log{S}$ 
relationship.
With a shallower $V_{c}-M_{acc}$ relation, the
black hole luminosities are spread over a narrower range,
and the simulated $\log{N}-\log{S}$
is much steeper. Remember that this assumed that there was
no population of quasars  associated with galaxies that 
were too small for us to resolve in the simulation (with halos $v_{c} \ll 100
\kms$). With a much higher resolution simulation, the situation could be
different.  However, for the bright end slope
to be reproduced, we would  still require a steep $V_{c}-M_{acc}$ relation.

\section{The Simulated Sky}

Looking at relatively thin slices of a CDM universe (e.g., 
Figure \ref{slice}) can
tell us about the morphology of the simulated X-ray
emission. However, to compare with observations, which give
us angular positions of X-ray photons on the plane of the sky, it is 
necessary to make simulated sky maps. In this way, we can see how much of
the filamentary structure of the IGM is preserved when projection is
taken into account, and whether any substructure in
clusters arises from background material. We can quantify the predictions
of our model with measurements of the angular clustering
on small scales. By making maps, we can also see how the AGN point sources
lie within the diffuse emission, and assess the effect of their different
redshift distributions.

\subsection{Map making}

Our simulation volume has a side length of $50 \hmpc$, but 
X-ray emission from AGN originates from a redshift of $z=2-3$, at a comoving
distance 50 times as far away or more.  To make maps, therefore, we
must replicate our simulation volume (including
a random change of viewing angle) many times. 
Scaramella, Cen \& Ostriker (1993) have made both
soft XRB and  Sunyaev-Zel'dovich maps using a similar
technique applied to lower resolution simulations of a CDM model 
(although with the advantage that they were able to use 
several different phase simulations). 
The same technique has been used by others for making maps
of the Sunyaev-Zel'dovich decrement: da Silva \etal (2000), 
Seljak, Burwell, \& Pen (2000),
and Springel et al.\ (2000),
 A similar idea has also been used for maps
of gravitational lensing distortions by White \& Hu (2000).
Unlike Sunyaev-Zel'dovich maps, the X-ray maps are subject to the 
inverse-square law, so that emission from low redshifts will dominate the
signal (particularly the clustering).
 This should mean that they will be less sensitive to 
the restrictions imposed by a small box, as at low $z$ the simulation
volume only subtends a small angle. We should, however, be careful
about interpreting our results, and we will be restricted to studying
clustering of the XRB on small scales 
(separations $\theta < 20$ arcmins). We leave study of large-scale 
structure in the XRB to future work, when larger simulations
become available.
Some tests of convergence, using different
box sizes and mass resolutions, have been carried out for 
the Sunyaev-Zel'dovich 
maps by Springel \etal (2000). Tests of numerical projection against
analytic clustering  (although not of X-ray emission)
in different contexts have been presented by,
e.g., White \& Hu (2000) and Croft \& Metzler (2000).

We have chosen a field of view $1\deg$ on a side to make our maps, 
and use $512^{2}$ pixels. The pixel size is therefore 7 arcsec,
roughly the same width as the FWHM of the {\it XMM-Newton} PSF,
and significantly coarser than that of the {\it Chandra} telescope.
We do not attempt to include  foreground 
emission or absorption in the maps, or to tailor them to any specific
instrument by including instrument noise or response functions.
Also, our maps do not include Poisson sampling noise from individual
photon statistics, and will include structure faint enough that 
it will be challenging to detect with current X-ray telescopes.

\begin{figure}[t]
\centering
\PSbox{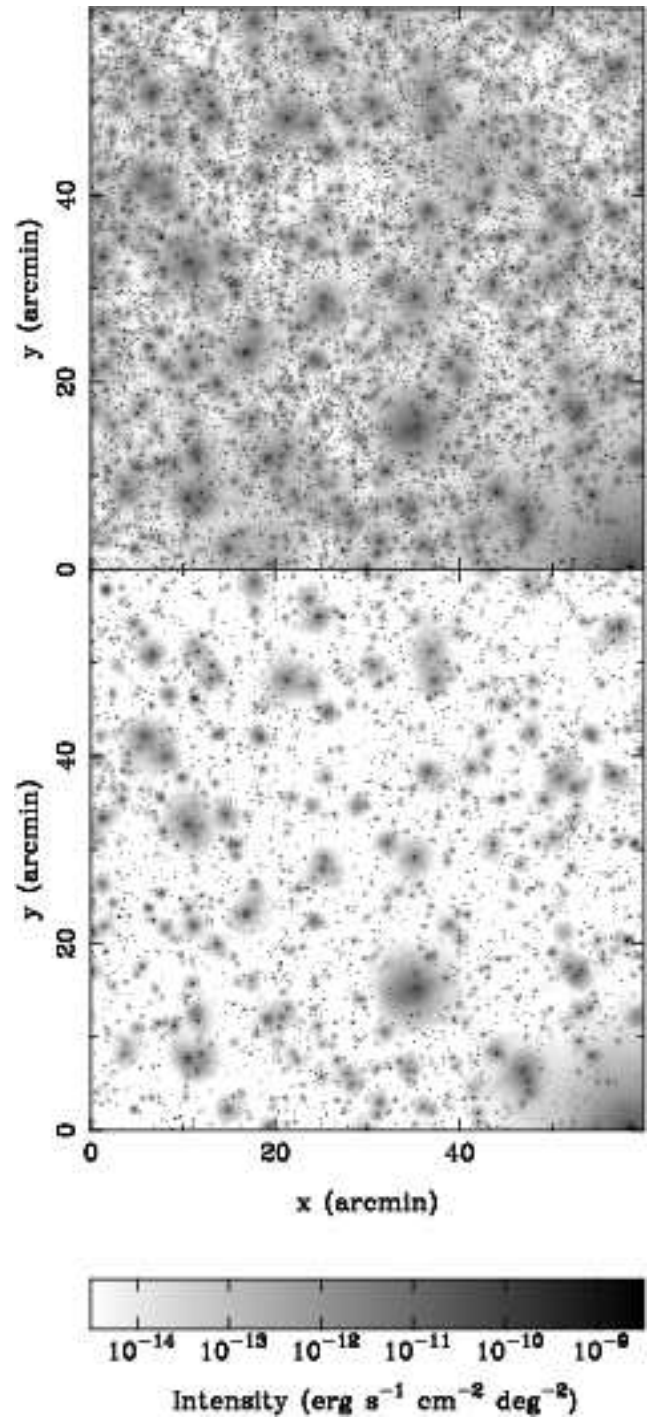 angle=-90 voffset=575 hoffset=-270 vscale=100 hscale=100}
{3.5in}{7.5in} 
\caption
{
Maps of diffuse emission in the soft band (top) and hard band (bottom).
\label{diffusemap}
}
\end{figure}

We make use of 27 simulation outputs, ranging in output 
redshift from $z=0$ to $z=6$, and spaced roughly logarithmically in redshift.
We have checked that there is no significant X-ray emission from beyond
$z=6$ (see e.g., Figure \ref{emvsz}). The comoving distance to $z=6$ is
5250 $\hmpc$, so that we replicate the box 105 times. We work in 
comoving coordinates, to take advantage of the fact that light rays travel
in straight lines in a flat universe. The boxes are stacked along the line of 
sight, using the output time closest to the redshift reached at
a given point. Each box is randomized, which involves a random
recentering of coordinates, and a one in two chance of a reflection
about each axis. We orient the box so that one of the axes (chosen at random)
lies along the line of sight, and also perform a random rotation about that
axis for simulation volumes close enough to the observer that the field of
view fits entirely within the box length divided by $\sqrt{2}$.
Above $z=1.4$, the simulation volume no longer subtends
an angle greater than $1 \deg$, so that we also periodically 
replicate the box across the line of sight. At $z=6$, the simulation
volume has an angular extent of $0.55 \deg$.

To assign the X-ray flux from the particles to the sky we again use the 
projected SPH smoothing kernel (equation \ref{spha}), to preserve
the high spatial resolution of the simulation in dense regions. For 
the AGN point sources,
we assign the fluxes to individual pixels. In the Figures showing
maps which include AGN, we have convolved the AGN with a finite
PSF (Gaussian with
7 arcsec FWHM) to make them stand out. In our statistical
analysis of the maps, however, we do not do this.
We calculate the luminosity distance for sources 
using the analytic approximation
given by Pen (1999b).

We make several different sets of maps, in both the soft band 
(0.5-2 keV) and the hard (2-10 keV).  To gauge the contribution of
different components, we make maps with either IGM gas that includes metals
(using the metallicity-density relationship of \S3), or without metals,
and other maps 
including only IGM gas with temperatures between $10^{5}$K and $10^{7}$K
(the WHIM).
As before, we make no distinction between the intracluster and intragroup 
medium and the IGM.
 We also make maps of the AGN component, and maps that are
a sum of the AGN and IGM maps, which will represent the overall
prediction for the XRB. Each set contains 20 maps, which are generated
by using different random seeds for the box randomization procedure.

As mentioned before, one aspect that we do not 
include explicitly in our maps is the 
absorption of X-rays by neutral hydrogen, either in 
our galaxy, or elsewhere. We have chosen the lower end of our soft band
($0.5$ keV) so that the predictions can be compared
to observed measurements without large corrections
for absorption. 
 Absorption in distant systems along the line of sight will
have too small a filling factor to have an effect on our maps.
As far as internal absorption in the AGN sources or their host galaxies
is concerned, we have normalized our simple model (\S4)
using the observed $\log N - \log S$ distribution, so that any such absorption
will be included implicitly, degenerate with the other parameters
that govern the model.

\subsection{Maps}

Example IGM maps (including the contribution of metals), for the soft and 
hard bands are shown in Figure \ref{diffusemap}. We can see that 
there is much more diffuse emission in the soft band, as we might expect,
with only clusters showing up in the hard band. Filaments are not 
as clearly seen in the soft band maps as in the 
single box plot (Figure \ref{slice})
owing to dilution by projection. There are some vaguely elongated
structures that might be filaments, particularly on the right side of 
the map. We examine them in more detail below. We can also see evidence
of substructure in the dense cluster regions.

\begin{figure}[t]
\centering
\PSbox{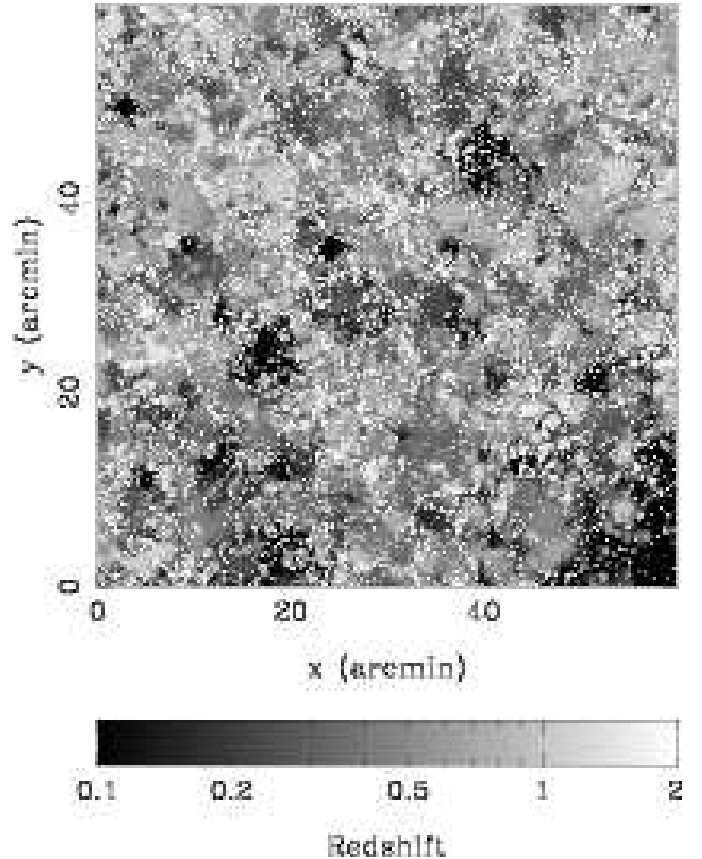 angle=-90 voffset=350 hoffset=-120 vscale=60 hscale=60}
{3.5in}{4.8in} 
\caption
{
Flux-weighted redshift (for diffuse emission in the soft band only).
\label{emwtz}
}
\end{figure}

To understand which features in Figure \ref{diffusemap}
 are coming from which 
redshift, we create a map of flux-weighted redshift. 
To make it, we assign the redshift of each particle multiplied
by its X-ray flux to the sky grid. We then divide out this map by 
the X-ray flux map. We have done this for the soft X-ray band (IGM only),
and we show it as Figure  \ref{emwtz}, which can be directly 
compared to the top panel of Figure \ref{diffusemap}. 
 By doing this comparison, we can see that what appears to be
a filamentary structure on the right-hand side of Figure \ref{diffusemap}
is composed of emission from several different redshifts ranging from 
$z \sim0.3-1.0$, and its appearance is a chance result of projection.
It is also evident that what appears to be ``substructure'' in the 
galaxy clusters (for example the nearby object in the bottom 
right hand corner) is also due to projection. The spatial pattern
of $z$ in Figure \ref{emwtz} is interesting in itself. 
The map is dominated by  many coherent, fairly round patches of extent 
$\sim 5$ arcmin, which are
clusters and groups at $z\sim0.2$ to $z\sim1$. These are ringed by 
more granular emission from the IGM at higher redshifts.
This was also pointed out recently
 by Voit, Evrard \& Bryan (2000), who predict that a large fraction of sky area
is covered by the superimposed virial regions of clusters at various 
distances from the observer, which complicates the search for filaments.
We will study the contribution to the total XRB arising 
from different redshifts and different components
in \S5.3 below.

\begin{figure}[t]
\centering
\PSbox{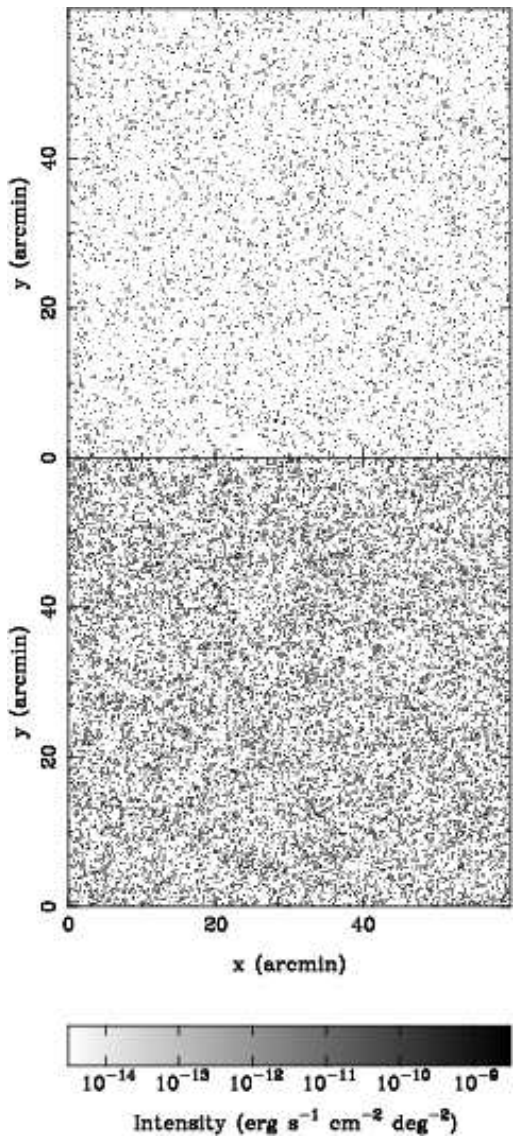 angle=-90 voffset=470 hoffset=-200 vscale=80 hscale=80}
{3.5in}{6.5in} 
\caption
{
Maps of point source emission in the soft band (top) and hard band (bottom).
To make the sources stand out, we have applied
a Gaussian PSF with a FWHM of
7 arcsec in this plot (not used in Figure \ref{diffusemap}).
\label{agnmap}
}
\end{figure}

In Figure \ref{agnmap}, we show maps of the AGN part of the XRB.
There is not anywhere near the amount of structure evident in the 
IGM maps. We will see later that the flux from AGN peaks at a higher
redshift, hence angular scales correspond to larger physical scales,
scales where there is not so much clustering. 
Our simulations cannot be used to study
clustering in the AGN component on larger scales, owing to the small 
box size. Also, field to field fluctuations
will be underestimated by missing power on scales greater than $50 
\hmpc$.

The $\log{N}-\log{S}$ plot (Figure \ref{lognlogs}) was made using  the sources
that contribute to maps like Figure \ref{agnmap}. As we expect from
the source counts,  there are many more sources visible in the hard band.
There are approximately 5000 sources for each 1 deg$^{2}$ in
the soft band with fluxes $>10^{-17} \ergscm$, and 16,000 sources
in the hard band. Below this flux limit, the source counts stay fairly flat,
because there are no small galaxies in the simulation to host
``mini-AGN.'' 
If such a population exists, the host galaxies  
would be below our resolution limit.
As we will see later, a new population of miniquasars
that contribute substantially to the XRB is unlikely in the context
of our model, as inclusion of X-rays from the diffuse IGM component  
already brings the total flux up to, or even slightly over, the observed level.
However, it is important to bear in mind that this assumes that
the IGM model is the correct one.
The contribution from the IGM could be made smaller
if we included stronger feedback, for example, which would leave room
for more faint AGN. Whether this is the case or not can be addressed both
by deeper imaging of the XRB, to look for point sources,
and by comparing model predictions with other properties of the XRB,
such as its clustering and spectrum.

The summed flux from the AGN which make up  Figure~\ref{agnmap} is
much greater than that from the IGM component, (see \S5.3 below),
although the great majority of sky pixels contain no AGN flux,
even at the relatively low resolution of our maps. 

\begin{figure*}
\centering
\PSbox{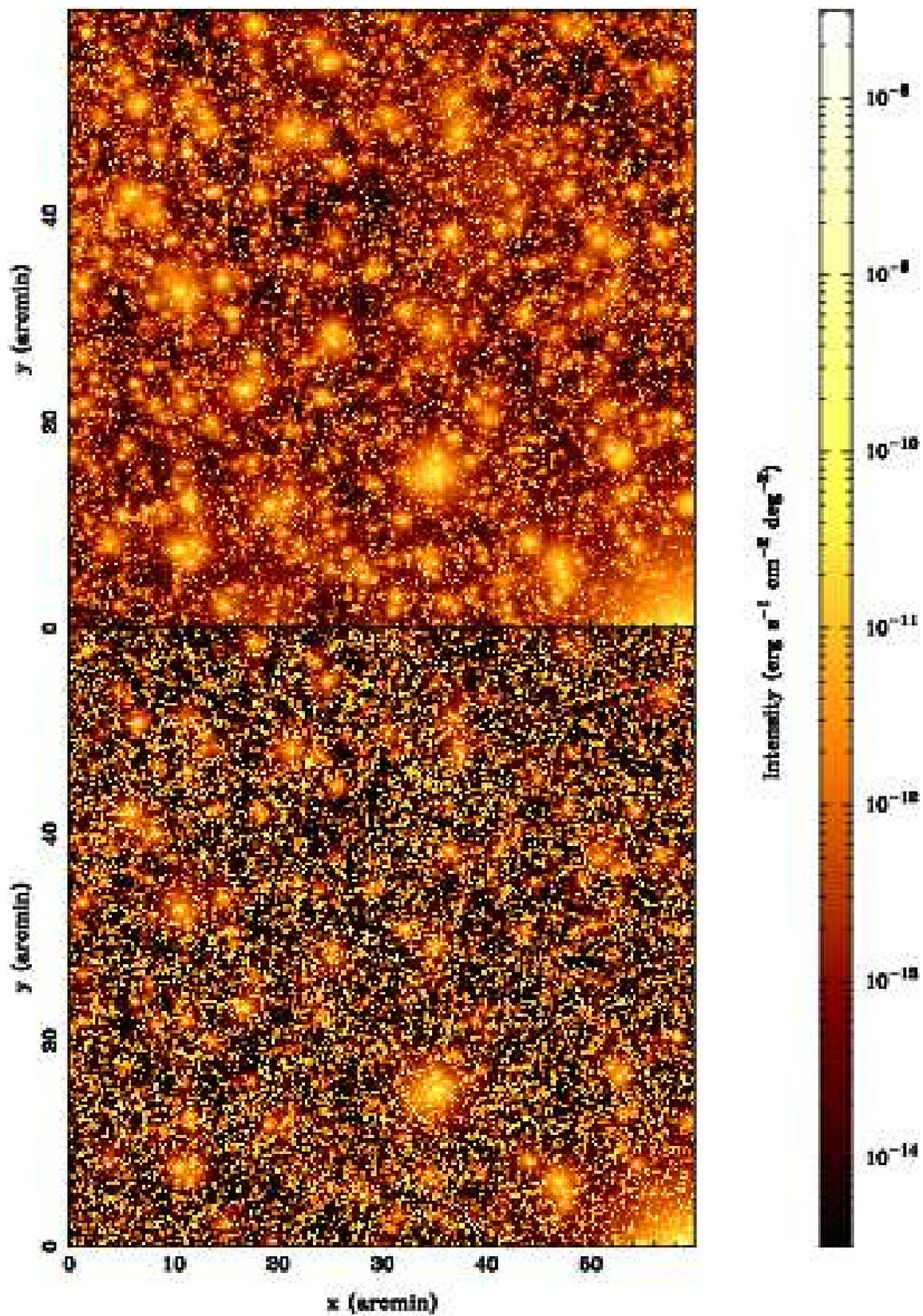 angle=-90 voffset=720 hoffset=-320 vscale=130 hscale=130}
{5.5in}{8.0in} 
\caption
{
Maps of total (diffuse + point source) emission in the soft 
band (top) and hard band (bottom).
\label{totmap}
}
\end{figure*}

The maps of the total XRB are shown in Figure \ref{totmap}, where we can 
compare the relative brightness of the AGN and IGM components. We have again 
used a 7 arcsec PSF for the AGN,
to make them more visible.
Even so, the AGN in the soft map are not very noticeable.
Of course, the main reason for this
is that the map contrast level is set to highlight features that are very faint.
The surface brightness in the soft band 
at the edges of clusters and groups, where
the emission trails off into the general IGM, is $< 10^{-12} \ergdeg$.
For comparison,  Scharf \etal (2000) have found evidence for a filament
with surface brightness $\simeq 6\times10^{-13} \ergdeg$
in a deep ROSAT PSPC field.

\begin{figure}[t]
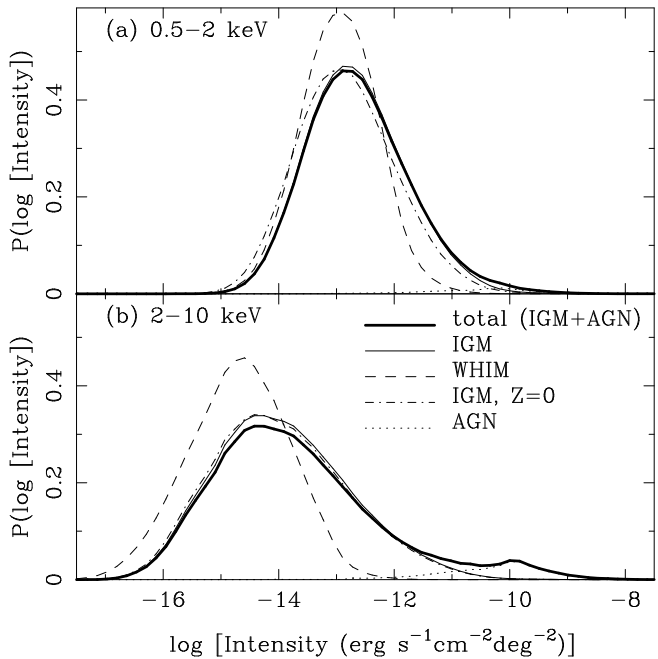

\centering
\PSbox{hist.ps angle=-90 voffset=290 hoffset=-60 vscale=48 hscale=48}
{3.5in}{3.7in} 
\caption
{
Probability distribution function of map pixel values for the soft (top)
and hard (bottom) bands.
\label{hist}
}
\end{figure}

The typical surface brightness levels of the map pixels can be seen from 
Figure \ref{hist}, where we plot the probability distribution
of pixel values (using 7 arcsec pixels for all components). We show
as separate curves the total XRB, and the PDFs computed
using IGM and AGN emission separately.
We also show histograms 
for the  WHIM on its own, and the IGM without the contribution of metals.
The median surface brightness of the total XRB in the soft
band is $1.7\times10^{-13} \ergdeg$,
compared to $1.0\times10^{-14} \ergdeg$ for the hard band, but the distribution
of pixel values is much wider for the latter.
We will see later that this manifests itself in a higher clustering
level for the hard IGM emission, although this is
canceled out in the clustering
of the total emission by the much brighter and more uniform AGN component. 
 The WHIM curve
is noticeably offset for the hard band, indicating as we would expect that 
this gas makes little contribution.
We will examine the average intensity and compare
to some observational determinations in \S5.3 below.

\begin{figure}[t]
\centering
\PSbox{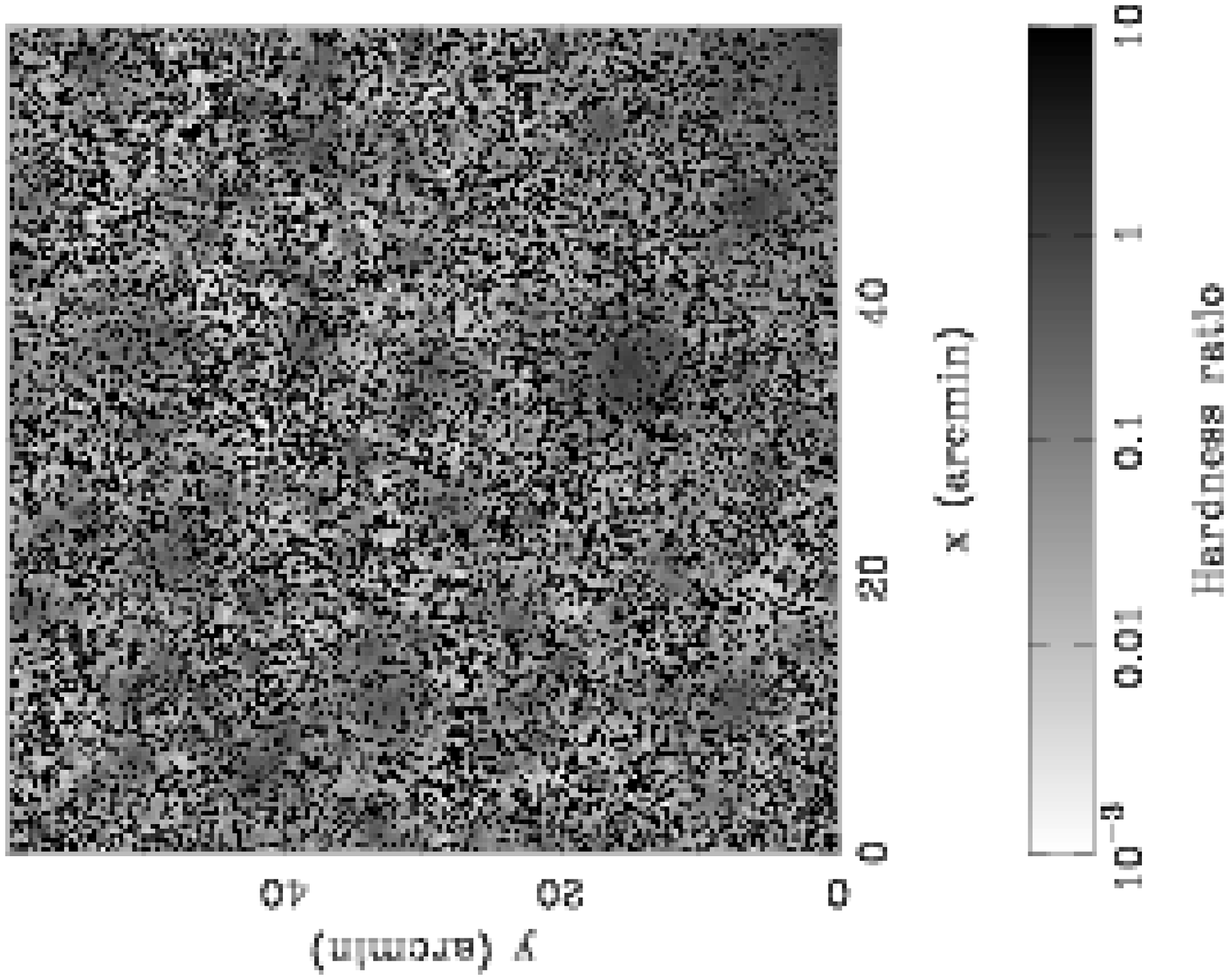 angle=-90 voffset=350 hoffset=-120 vscale=60 hscale=60}
{3.5in}{4.8in} 
\caption
{
Hardness ratio (hard band/soft band flux), for the total
XRB map (AGN+IGM, Figure \ref{totmap}).
\label{hr}
}
\end{figure}

We also plot
spectral information by mapping the hardness ratio (HR). We 
define this quantity to be the hard band flux divided by the soft band flux,
and show the
results in  Figure \ref{hr}. From the plot, it is obvious that the 
AGN have 
much harder spectra than the IGM  (even harder than gas in the center
of clusters). 
Because of the way the relative sizes of the soft and hard bandpasses
are defined, none of the IGM emission has a HR $ \gtrsim 0.5$.
A small dynamic range in HR covers the difference
between cluster centers and the more diffuse material in the outskirts.
This latter material, therefore, shows up more dramatically in the HR map than
in the maps of Figure \ref{diffusemap}. As we expect, the majority of
the sky area is soft, with a HR of 0.01 and below,
even though the total flux is higher in the hard band
(due to the AGN). Because we are dividing one map by another, 
very deep integrations would be needed to recover any detail in the 
faint, soft regions. Averaging over fairly large areas of the map
should enable us to recover some information about the 
temperature and density distribution of the diffuse IGM.

One subtle effect that might be worth exploring is the way clumps of
IGM at different redshifts have different hardness ratios
because of the shifting bandpass. The spectral 
information present may give hints as to the redshift distribution
of the diffuse emission, although, of course, this will be convolved
with differences in HR due to temperature variations.

\subsection{Mean total intensity}

\begin{table}
\centering
\caption[totalem]{\label{totalem}
Mean XRB intensity. \textnormal{The error quoted is the 
dispersion in 1 deg fields,
and not the error on the mean.}}  
\begin{tabular}{ccc}
\hline &&\\
Component   &  0.5-2 keV band  & 2-10 keV band \\
            & $\ergdeg$         & $\ergdeg$ \\
\hline &&\\
 Total &  $(6.58 \pm 0.88) \times 10^{-12}$  &
 $ (2.60 \pm 0.16) \times 10^{-11}$ \\
 IGM &  $(2.29 \pm 0.43) \times 10^{-12}$  &
 $ (9.67 \pm 3.90) \times 10^{-13}$ \\
 IGM Z=0 &  $(1.74 \pm 0.34) \times 10^{-12}$  &
 $ (8.99\pm 3.80) \times 10^{-13}$ \\
 WHIM &  $(4.15 \pm 0.55) \times 10 ^{-13} $ &
 $ (1.47 \pm 0.41) \times 10^{-14}$ \\
 AGN &  $(4.29 \pm 0.76) \times 10^{-12}$  &
 $ (2.50 \pm 0.16) \times 10^{-11}$ \\
\hline &&\\
\end{tabular}
\end{table}

In Table \ref{totalem}, 
 we give the  averaged total intensity in the XRB from 
our 20 maps. 
We also give results for the different components (AGN, WHIM, IGM, IGM with 
zero metallicity) taken separately, as well as the dispersion in 
results taken from the scatter in the 20, one degree fields.

The AGN are dominant in both bands, accounting for $65\%$ of the
soft-band emission and $96\%$ in the hard-band. 
Of the IGM fraction of the XRB,
line emission from metals  accounts for
$24 \%$ of the flux. In the hard band, as expected,
the Bremsstrahlung contribution is higher, so that only $7 \%$
is produced by the metals. 
The WHIM (intergalactic
gas with temperatures between $10^5$ K and $10^7$ K) is responsible
for $18 \%$ of the soft-band flux coming from the IGM.

If we compare to observations, we are most interested in the
situation in the soft-band.
Our total soft band XRB flux is close to the {\it Chandra} results
of Mushotzky \etal (2000), who find $6.0 \pm 1.5 \ergdeg$
for $0.5-2$ keV, 
 compared to our $6.58 \ergdeg$.

For the percentage contributed by AGN, we compare to the
ROSAT Lockman hole observations of Hasinger \etal (1998)
and Schmidt \etal (1998), 
as well as the sources in Mushotzky \etal (2000)
for objects at fainter flux limits.
 The ROSAT work  resolved
$68 - 81 \%$ of the soft XRB above a point source flux limit of 
$10^{-15} \ergscm$
into AGN. Mushotzky \etal (2000) find that  $6-13\%$ 
more of the XRB is resolved when they push to a
flux limit that is $\sim5$ times fainter.
Combining these results, from $74-94\%$ of the soft XRB flux
is observed to be in discrete sources.
These figures are, however, not for the $0.5-2$ keV band
but for a band from $1-2$ keV, which we would 
expect to be slightly more dominated by AGN.
Accounting for this slightly different band raises the AGN
fraction in our simulation, and we predict an AGN fraction $72 \%$,
which is $\sim 1\sigma$ less
than the observational estimate.  We conclude that our simulation
slightly overpredicts the diffuse component of the XRB, 
but given the observational and numerical uncertainties the level
of agreement is quite encouraging.

In the hard band, a large angle survey by 
Marshall \etal (1980) using HEAO1 A2 data found an average intensity of
$1.6-2.3 \times 10^{-11} \ergdeg$, compared to our 
total intensity of $2.6 \times 10^{-11} \ergdeg$.
Mushotzky \etal (2000), with the high resolution of 
the {\it Chandra} satellite,
were able to resolve $56-81\%$ of the hard-band XRB sources, above a flux
limit of $2.5 \times 10^{-15} \ergscm$. Given that we chose the
parameters for our simulated AGN to
reproduce the Mushotzky \etal $\log{N}-\log{S}$ relation, we
expect the total flux from our AGN to be
consistent with the AGN data in the hard band.  Because we have
 sources in our simulation
that are fainter than the Mushotzky \etal limit,  even more 
($96 \%$) of our 2-10 keV XRB comes from AGN,
with only a small hot cluster contribution making up the remainder.

If we consider radiation at lower energies than our soft band,
galactic absorption and coronal 
emission become extremely important.  We therefore do not plot any maps of
emission in a softer band, as its clustering
is likely to be too difficult to measure observationally.
We have, however, computed the mean intensity due to the IGM in the band 
centered on $0.25$ keV ($0.1-0.4$ keV),
and find a value of $1.9 \times 10^{-12} \ergdeg$,
which is $13 \kevkev$ in the units used by the observational
papers. This relatively low value is
as we might expect given the study of
CO99, who find that an energy of $0.7$ keV, an
 intensity of $7 \kevkev$ is produced by   
warm-hot intergalactic gas in their simulation.

The observational papers report on the use of
shadowing by foreground neutral hydrogen
to estimate the extragalactic
flux in the very soft X-rays. Cui \etal (1996) found $95\%$ lower and
upper limits at $0.25$ keV energies of $32\kevkev$ and $65 \kevkev$.
They state that at least 30 $\kevkev$ has been resolved into
discrete sources, so that $ \lesssim 15 \kevkev$ must be due to the IGM. 
Warwick \& Roberts (1998), in a review, give estimates
of $20-35 \kevkev$ for the total intensity, from
which Wu, Fabian \& Nulsen (1999) deduce that $< 4 \kevkev$
can be due to a diffuse component. 
At a slightly higher energy, 0.7 keV, Wang \& Ye (1996)
find an intensity of $28 \kevkev$, with $<14 \kevkev$ expected to
be due to discrete sources.
 As with emission in the 0.5--2 keV band, the simulation therefore
gives results for diffuse emission 
in a softer band that are close to, or maybe
slightly higher than, the observations.

\subsection{Intensity versus redshift}

\begin{figure}[t]
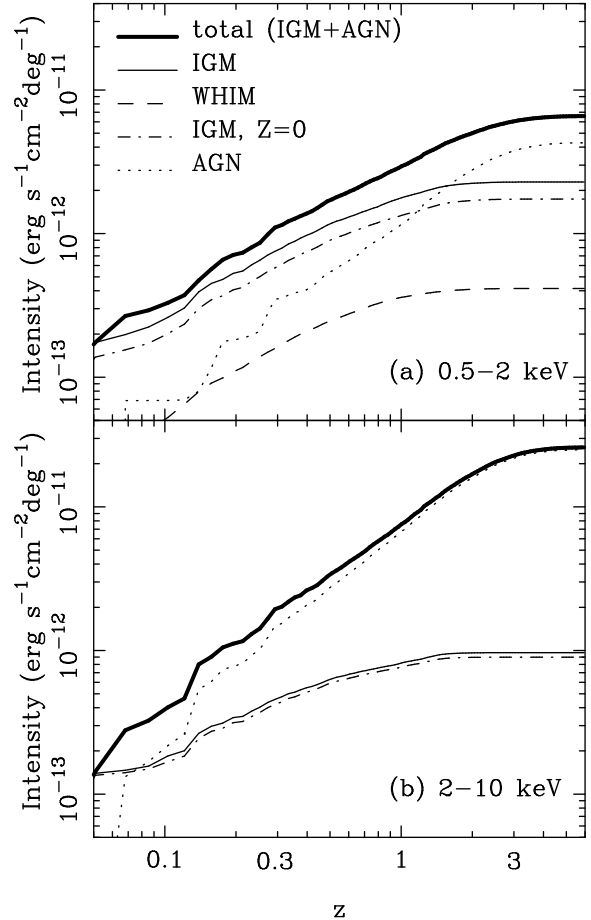

\centering
\PSbox{emvsz.ps angle=-90 voffset=400 hoffset=-155 vscale=70 hscale=70}
{3.5in}{4.8in} 
\caption
{
Cumulative flux as a function of redshift in the soft (top) and hard bands
(bottom). We show the contribution of AGN, diffuse gas, diffuse gas
without the contribution of metal line emission,
and gas between $10^{5}$ and $10^{7}$K, as lines of different types.
\label{emvsz}
}
\end{figure}

In Figure \ref{emvsz}, we show the intensity in the XRB emitted below
a given redshift, for the different components. The values quoted in
Table \ref{totalem} correspond to the intensity reached by $z=6$, on the
right side of the plot. 

In the soft band, the IGM emission dominates for $z \lesssim 1.5$, above which 
it makes no significant contribution.
The median redshift of total emission in this band is $z=1.1$, and for 
the IGM component it is $z=0.45$.
 If we  consider the differential amount
of emission per unit redshift (not plotted), the IGM XRB
intensity peaks at $z \sim0.2$ and then declines fairly slowly, reaching
$\sim \frac{1}{3}$ of its peak value at $z=1$. Because the IGM dominates 
at low $z$, there will be significant
clustering due to it in the soft band, which we have already seen in
the maps, and which we will quantify below.
Bandpass shifting affects the fraction of WHIM emission in the IGM,
which declines slightly as we move to higher redshift, as seen in 
Figure \ref{emvsz}.

\section{Angular Clustering}

\subsection{The autocorrelation function}

We quantify the amount of clustering on different angular scales
in our maps using the angular autocorrelation function,
$w(\theta)$. Our estimator is 
\begin{equation}
w(\theta)=\langle \delta_{X}({\bf r}) 
\delta_{X}({\bf r} + {\bf \theta}) \rangle,
\label{wthetaestimator}
\end{equation}
where ${\bf r}$ is the angular coordinate of the center of a map pixel,
and $\delta_{X}({\bf r})=[I_{X}({\bf r})/\langle I \rangle]-1$, where 
 $I_{X}({\bf r})$ is 
the X-ray intensity in that pixel.
 We apply the estimator
to the individual 7 arcsec pixels for small $\theta$, but regrid
the maps to coarser pixels on larger scales to make the calculation faster.
We have checked that both estimates match in an overlap 
region as expected. 

\begin{figure}[t]
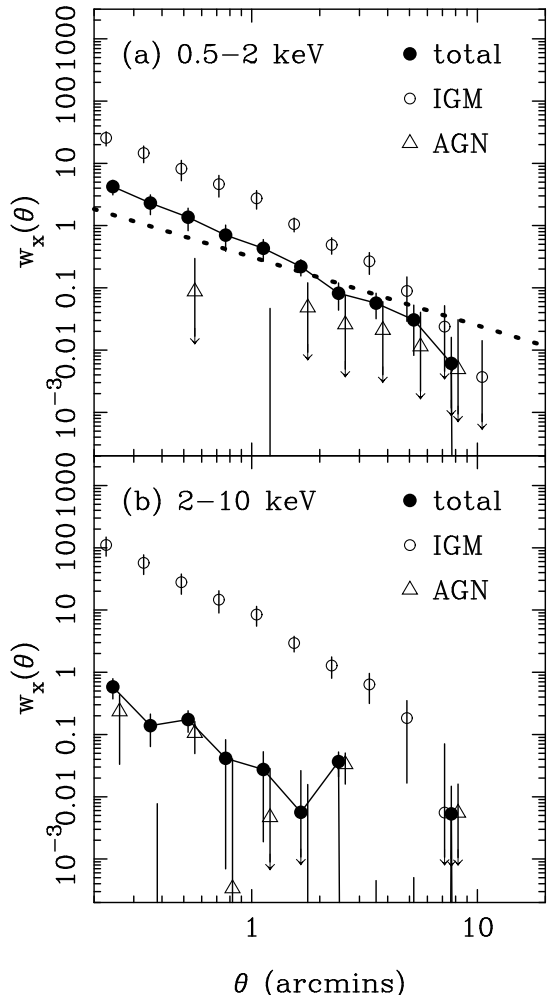

\centering
\PSbox{wtheta.ps angle=-90 voffset=430 hoffset=-160 vscale=75 hscale=75}
{3.5in}{5.2in} 
\caption
{
The angular correlation function of the XRB. We show results calculated
from the IGM and the AGN maps separately, as well as for the summed map.
The error bars are the error on the mean calculated from the scatter
between measurements from 20 different maps each of area
1 deg$^{2}$. The dotted line is a power-law fit to the ROSAT All-Sky survey
results of Soltan \etal (1999). This fit was measured 
from data on angular separations from
20 arcmin to 20 deg., so that the range of scales we show here represents an 
extrapolation.
\label{wtheta}
}
\end{figure}

The results are shown in Figure \ref{wtheta}. In the soft band, 
$w(\theta)$ for the IGM alone is roughly a power law
with an index of $-1.4$, and falls away on large scales. As with all such 
 measurements, we should be careful about interpreting the large-scale points,
because they will be compromised by the small box size. 
For example, 10 arcmin corresponds
to $\sim 4$ comoving $\hmpc$ at $z=0.45$, the median redshift 
of the IGM emission.
Scaramella, Cen \& Ostriker (1993), in their study of $w(\theta)$
measured from simulated XRB sky maps also found a relatively steep dependence
on angle. Their results were best fit by an exponential on scales
 $\theta < 10 $ arcmin, after which $w(\theta)$ became negative.

The AGN
component in our maps
yields hardly any signal, so the clustering of the total XRB
is basically that of the IGM diluted by a factor
of $0.35^{2}$ (where 0.35 is the ratio of total intensities in
the IGM and the total XRB).
Measurements of strong clustering in the XRB on these scales, therefore,
point towards the existence of baryons in the diffuse IGM,
and in galaxy clusters
(Blanchard \etal 1992; Soltan \etal 1996).

Published observational data is available on slightly larger scales
than those we are able to measure here. Soltan \etal (1999)
have estimated $w(\theta)$ from the soft band ROSAT All-Sky Survey
for $\theta > 20$ arcmin, finding approximately a power law of slope $-1.1$.
In making these measurements, Soltan et al.\
removed the galactic component of the XRB, which introduces 
major uncertainties related to the possible
presence of residual fluctuations.
They made no attempt to remove distinct sources, so that we can
compare directly with our ``total'' results, by 
 extrapolating their results to smaller scales (the dotted line in 
Figure \ref{wtheta}). We find reasonable agreement, about as good as can be 
expected, given that it is an extrapolation, and that there are uncertainties
on both sides. If only the AGN component contributed
to the clustering, the simulation would lie at least a factor of five too low.
On the larger scales actually probed by the  Soltan $\etal$ measurement,
it seems likely that AGN come to dominate the clustering, given that 
the slope of the (very noisy) AGN $w(\theta)$ appears to be flatter.
Larger simulations are needed to verify this, however, since
the finite size of the simulation volume tends to suppress both
IGM and galaxy correlations on large scales.
A semi-analytic model of the XRB clustering due to AGN 
by Haiman \& Hui (2000) finds a similar, low amplitude to our
AGN-only results.

In the hard band, the IGM component has a higher clustering level 
than in the soft X-rays, because only the 
highly clustered central
regions of relatively bright clusters are being sampled. However, because
the AGN fraction of the XRB is much higher ($96 \%$),
this IGM clustering is suppressed very strongly when we consider the 
total XRB. The AGN seem to have detectable clustering on
smaller scales than in the soft band, probably because of the higher space
density of hard band sources. The total $w(\theta)$ is not much higher
than the AGN $w(\theta)$ value.  Note that the clustering properties of the
galaxies in the simulations are consistent with observations both at low
and high redshift (Dav\'e \etal 2001), so our inferred clustering of the AGN
should be realistic.

\subsection{Cross-correlation of the XRB with nearby galaxies}

We have seen that much of the IGM emission 
in the simulated maps occurs at low redshift. One way to 
check that this is also true in the observations is to cross-correlate
the XRB intensity with low redshift galaxies. 
Such an analysis was carried out by Soltan \etal (1997), using the
ROSAT All Sky Survey and bright local galaxy catalogs. The
angular cross-correlation function was measured
on larger scales than we are able to simulate, and a significant signal
was measured.  
Here we carry out an analysis of the cross-correlations 
on small scales, with a deeper simulated galaxy survey.

In \S4, galaxies were picked from the simulation and a fraction of them
 denoted to have AGN. We will use the same initial set of galaxies here but
restrict ourselves to a complete sample uncompromised by numerical resolution,
i.e. ones that contain 60 or more particles.
We use these galaxies to make a low redshift flux-limited sample by assigning 
a luminosity to each galaxy proportional to its stellar mass,
and then applying a lower flux limit so that the mean redshift
of all the remaining galaxies is 0.2. We do the same for 20 
different surveys of one degree fields, generated using the same
randomization parameters as the X-ray maps (\S5.2). The same flux limit
is used for all 20 surveys, which contain on average $\sim 500$ galaxies each.

\begin{figure}[t]
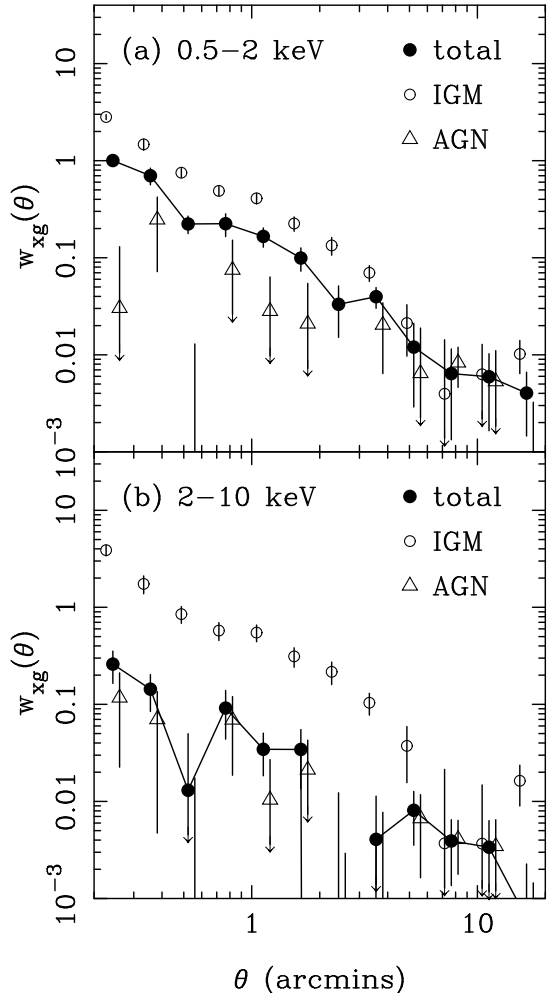

\centering
\PSbox{wxg.ps angle=-90 voffset=430 hoffset=-160 vscale=75 hscale=75}
{3.5in}{5.2in} 
\caption
{
The angular cross-correlation function of galaxies and the XRB.
\label{wxg}
}
\end{figure}

Our estimator for the cross-correlation function, $w_{Xg}(\theta)$,
is analogous to that used for the autocorrelation function
(equation~\ref{wthetaestimator}), so that we have 
\begin{equation}
w_{Xg}(\theta)=\langle \delta_{X}({\bf r}) 
\delta_{g}({\bf r} + {\bf \theta}) \rangle,
\end{equation}
where $\delta_{g}({\bf r})=[\rho_{g}({\bf r})/\langle \rho_{g} \rangle] -1$,
and  $\rho_{g}({\bf r})$ is 
the galaxy surface density.

We show the results in Figure~\ref{wxg}.  As with 
$w(\theta)$, there is a stronger signal in the soft band,
despite the fact that the mean galaxy redshift is
significantly lower than the median redshift of the emission.  
Observing the cross-correlation of the XRB with galaxy
samples of different mean redshift would be 
one way of studying the redshift distribution of the IGM emission.

\section{Summary and Discussion}

We have examined the distribution of X-ray emission  
from intergalactic gas in a hydrodynamic 
simulation of a CDM model dominated by a cosmological constant.
We also selected
a certain fraction of galaxies in the simulation
to have active nuclei, and calculated the X-ray
emission from these AGN using a simple model with free parameters
tuned to fit observed source counts. 
We projected the simulation volume, including
dimming and redshift evolution to make artificial sky maps.

Some of our main conclusions drawn 
from an analysis of these maps are that:

(a) The mean total XRB intensity in the soft ($0.5-2$ keV)
band, is $6.58 \times 10^{-12} \ergdeg$, with
$35\%$  being
generated by the diffuse IGM: clusters, filaments and groups.
$50 \%$ of this IGM emission comes from below a redshift of $z=0.45$,
and the XRB coming from $z>1.5$ is almost entirely due to AGN.
The mean intensity is in reasonable agreement
 with the observational value ($6.0 \pm 1.5 \time 10^{-12} \ergdeg$).
The fraction of the total due to the IGM is slightly overpredicted
by the simulations, as observationally less than
$\sim 25 \%$ has yet to be resolved into discrete sources.

(b) In the hard ($2-10$ keV) band, only $4 \%$ of the emission comes
from the IGM, with the median redshift of
XRB emission being $z=1.5$.
The total intensity is in agreement with observations,
as is the IGM fraction ($\lesssim 40 \%$ observationally).

(c) Because of projection,
it is difficult to see any evidence for the
filamentary emission, although this structure
is obvious in thin slices taken 
through the simulation. Diffuse emission from relatively low
densities does, however, cover much of the sky, and the median
soft-band intensity of map pixels is  $1.7\times10^{-13} \ergdeg$.

(d) The clustering measured from the maps is strong on the 
small scales we are able to
simulate ($<20$ arcmin), and fairly close to an extrapolation
of ROSAT All Sky Survey results to these scales.

(e) The angular
cross-correlation of the simulated X-ray sky with simulated bright
galaxy catalogs also yields a strong signal.

The procedure that we have presented here for making XRB predictions
probably represents the best that can be 
achieved using the hydrodynamic simulation available to us at present.
It will be subject to many possible improvements once more and
larger simulations can be used. One aspect that should be tested 
if we are to achieve more precise results is the effect of box replication
on the maps. We are using the same simulation, repeated along the line of sight
in the maps we have made here. This should not have any adverse effect as
long as clustering on the box scale is linear and as long as
the contribution of matter clustering on scales larger than the box is
not significant. Our present box of size $50 \hmpc$ is certainly smaller
than one would like, although the non-linearities on the
scale of the box evident in Figure \ref{slice} are not necessarily severe,
because only a tiny fraction of the box at these
low redshifts contributes to the maps. In the future it will
become possible to run convergence tests of box size on the results,
 and also use different phase simulations along the line of sight. 
Our results on the small scales we have concentrated on here should
not change significantly. For example, Springel et al.\ (2000) give evidence
that Sunyaev-Zel'dovich map fluctuations have converged in a
similar box size to ours. 

Another reason to use a larger simulation volume is so that we do not
under represent rare objects such as rich galaxy clusters, which 
will account for a large fraction of the emission. 
Because of this, the estimate of the XRB intensity from our
relatively small simulation is more likely to be an underestimate than
an overestimate.
The maps we have made in this paper are not useful for studying the
high luminosity tail of the emission, or individual large clusters.

Higher mass resolution would help us to 
represent the separate hot and cold gas phases more faithfully.
This would improve the situation that results
when hot IGM particles are close to clumps of
cold galactic gas and are therefore assigned unrealistically high densities. 
We have investigated this problem and its effects on the
X-ray emission in detail in Appendix A.
The solution we have adopted with the present simulation is
to separate the cold dense particles from 
the hot IGM after the simulation was run, and recalculate the densities,
before calculating the X-ray emission. 
We have tested this technique in Appendix A to show that it is
robust and gives the correct result.
In the future, a different approach to calculating  the SPH densities
could help, and if it were incorporated into the running of the simulation 
it may have some effect on galaxy formation by removing some artificially
enhanced cooling.

Simulations with higher mass resolution would also be very useful 
to gauge the effect on the total X-ray flux. With finite 
resolution, one will miss small objects, and
smooth out the cores of larger ones, which will yield 
an underestimate of the mean X-ray emissivity.
That this should not be severe problem 
is indicated by the study of D00, who show that a substantial
fraction of the X-ray emitting gas remains diffuse, with little
variation between  simulations that use widely varying
numerical techniques and have very different mass and spatial
resolutions.
We have also carried out a modest resolution test, described in \S A.4 below.
Although we have only one relatively large group at high resolution,
we find that its luminosity is only $75\%$
larger than the mean for comparable low resolution groups,
even though the mass resolution is 8 times better. Lower mass groups
make very little contribution to the average XRB.

The question of how much differing levels of feedback from galaxies affects
the IGM X-ray emission is something we have not investigated
in this paper. It has been shown 
in other simulations (e.g., Pierre et al.\ 2000;
Springel et al.\ 2000) that non-gravitational heating  
can potentially drive the IGM out of small groups of galaxies and
add to the more diffuse component. Some sort of strong 
feedback is also often invoked as being necessary for the observed
cluster temperature-luminosity relationship to be reproduced
(e.g., Kaiser 1991; Metzler \& Evrard 1994; Bryan \& Norman 1998).
The simulation we use here includes star formation, and a mechanism for 
returning
the energy of supernovae as thermal feedback to the IGM
(see Katz \etal 1996 for details). Because of the high spatial
resolution of the
SPH code, this feedback energy is deposited in very
dense regions, and it is quickly
radiated away, having only a minor effect on the IGM dynamics. Despite
the modest role of feedback,
a large fraction of gas with soft XRB temperatures is diffuse,
having not yet fallen into virialized structures, and is heated by shocks.
As shown by D00, this component
contains a large fraction of the baryons ($\sim 30\%$ by mass
are in this WHIM, with temperatures between $10^5$ and $10^7$ K).

In our model, as in that of
CO99, much of the very soft XRB is produced by this diffuse
WHIM gas, rather than by galaxy groups and halos.
CO99 showed that in their simulation, one quarter (7 $\kevkev$) of the 
XRB at 0.7 keV is produced by this gas.
  We have calculated
the mean intensity in our simulations in a band centered around 0.25
keV, and found it to be 13 $\kevkev$.  Observationally, different
analyses (Cui \etal 1996, Wang \& Ye 1996, Warwick \& Roberts 1998)
find that after subtracting the AGN contribution in the very soft
X-rays, at most $\sim 4-15 \kevkev$ can be in an IGM component.  This
is fairly close to our simulation predictions, although probably a bit
lower.  

Several authors have recently argued that supernova feedback of about
1 keV per baryon is required to prevent the gas in galaxy halos and
groups from over-producing the diffuse soft XRB.  Pen (1999) reached
an estimate of $230 \kevkev$ at 0.25 keV, using a hydrodynamical
simulation without cooling and assuming all the baryons were gaseous
with an emission-weighted temperature of $1.5 \times 10^6 \, {\rm K}$, typical
of galactic halos. Wu, Fabian, \& Nulsen (1999) used a semi-analytic
model to estimate a background of 35 to $60 \kevkev$ at 0.25 keV, in
the absence of supernova feedback.  This calculation allowed the gas
temperature to vary as a function of group circular velocity,
an improvement on the Pen (1999) calculation; the
resulting background was dominated by groups of galaxies.

Our simulations, by contrast, demonstrate that extreme supernova
feedback is not required to match the observed intensity
of 0.25 keV emission. 
 There are
several reasons why our conclusions differ from those of Pen and Wu \etal
(see also the discussion in Dav\'{e} \etal 2000).  First, as
mentioned above, only 30\% of the gas is in the WHIM.  As Bryan (2000)
also points out, the efficiency of forming galaxies is higher in
groups than in clusters, which removes much of the gas 
in these lower mass systems.
The papers cited above assumed that essentially all of the
baryons in any virialized halo were in a hot X-ray emitting medium.  
Second, the gas
in these papers is assumed to be isothermal in the potential of a
cuspy dark matter halo, so that most of the emission is radiated at
the center of the halo.  In our simulation, we find instead that the
X-ray-emitting gas typically has a core with a radius of several
hundred kpc (see Figure 7 of Dav\'{e} et al.\ 2000, which is based on
this simulation).  The strong emission at the center is thus
eliminated.  This core may be somewhat influenced by our numerical
resolution, but the analytic model by Bryan (2000) also finds such a
core.  Third, in our simulations
much of the energy released in gravitational clustering emerges in the
form of atomic lines, principally Ly$\alpha$, rather than in X-rays
(Fardal \etal 2000).  As our XRB predictions are slightly higher than the
observational limits, it is possible that a modest increase of
feedback in the simulations would improve the agreement.  We will
report on a study of the X-ray luminosity function of galaxy groups,
which are most sensitive to feedback effects, in future work.

On the observational side, maps of extragalactic soft X-ray
emission are extremely difficult to make, mainly
because of emission and absorption
by our galaxy. The ROSAT All Sky Survey is 
dominated by galactic thermal emission over much of the sky,
 and even in regions
close to the galactic poles, there is uncertainty in the
contribution of this component to the measured X-ray fluctuations
(see, e.g., Soltan \etal 1996). It is possible that
realistic theoretical maps such as those we have presented here can play a role
in testing the procedures used to remove galactic foreground contamination,
just as they do in the study of the Cosmic Microwave Background
(e.g., Bouchet \& Gispert 1999).

Ultimately, one can hope to study cosmology with the small scale
structure of the XRB.  
The contribution from various species of AGN is certainly 
complicated, but because the AGN emission is widely spread in redshift,
the clustering of the soft XRB
seems to be dominated (at least in the model we have studied)
by the IGM. The physical processes involved in the emission are relatively
simple, although the question of feedback and the effect it
will have on small scale structure must be studied.
The new X-ray satellites, {\it Chandra} and {\it XMM-Newton},
should soon measure clustering accurately on the scales for which we
have made predictions in this paper.
Among the many further analyses that
can be made are cross-correlations of the XRB with Sunyaev-Zel'dovich
and weak-lensing maps, as well as measuring the cross-correlation
with galaxies as function of redshift. We also hope that simulated skymaps
such as ours will help advance methods for detecting emission
from the WHIM and thus help to complete the census of local baryons.

\bigskip
\acknowledgments
We thank Volker Springel, Chris Metzler and Max Tegmark for useful discussions.
This work was supported by
NASA Astrophysical Theory Grants NAG5-3820, NAG5-3922, and  NAG5-3111,
by NASA Long-Term Space Astrophysics Grant NAG5-3525, and by the NSF under
grants ASC93-18185, ACI96-19019, and AST-9802568.
TDM is supported by NASA through Chandra Fellowship
grant number PF8-10005 awarded by the Chandra Science
Center, which is operated by the Smithsonian Astrophysical
Observatory for NASA under contract NAS8-39073.
The simulations were performed at the San Diego Supercomputer Center,
NCSA, and the NASA/Ames Research Center.

\appendix

\section{Allowing for a two-phase medium}

\subsection{Introduction}

The simulation includes the cooling processes that 
transform collapsing dense gas into cold galactic clumps that can then 
form stars. This galactic gas is often in close proximity to
intracluster gas, which is in a completely different phase,
at least $\sim100$ times less dense and $\sim 100$ times hotter or
more. The SPH density estimate we use (Hernquist \& Katz 1989)
is evaluated from the positions of the 32 nearest neighbours of
each particle, regardless of their temperature. 
With very high mass resolution, there should be no problems with resolving
the distinct gas phases. However, in practice we have a limited number
of particles, so that the standard density estimate for some
hot intracluster gas that is close to galaxies will
include cold, much denser particles, which are in a different phase.
This results in a spuriously high density estimate for these hot
particles, something that is especially problematic when calculating
their X-ray emission. For our present purposes, this problem
can be easily mitigated by separating the hot and cold dense gas phases 
before evaluating the density, after the simulation has been run.
In the same way that the baryonic
material in star particles does not contribute to gas
particle density estimates, we remove the cold, collapsed gas
from the calculation of the density of hot gas.

In this Appendix we will illustrate the effect of this correction on the
X-ray luminosities of particles, and test that it is robust by varying the
thresholds we use for excluding particles. We will also
carry out a simple test involving an isothermal sphere with a cold
gas clump at the center to show that the effect can be reproduced in a
controlled situation. Finally, we compare the X-ray luminosity of
a galaxy group in a smaller, higher mass resolution simulation
to similar groups in our large simulation, to
check that separating the gas phases has the expected smaller effect
on the X-ray luminosity at higher resolution.

We note that different solutions for dealing with the effect of
two gas phases on galaxy formation in  SPH simulations have been proposed
by Pearce \etal (1999), and Ritchie \& Thomas (2000).
A detailed study of the effect on galaxies should be carried out by running
simulations with a different method for estimating the density
(e.g., a direct solution of the continuity equation [V. Springel,
private communication]), and is beyond the scope of this paper.

\subsection{Two-phase density estimates}

We recalculate the density estimates for hot particles
in the simulation by excluding cold, dense particles from the
kernel estimation. We define hot particles
to be those with $T>10^5$K  and set the threshold for exclusion to 
be $T< 10^{4.5}$K and $\rho > 1000$. The effect of varying these thresholds
is negligible (see below).

\begin{figure}[t]
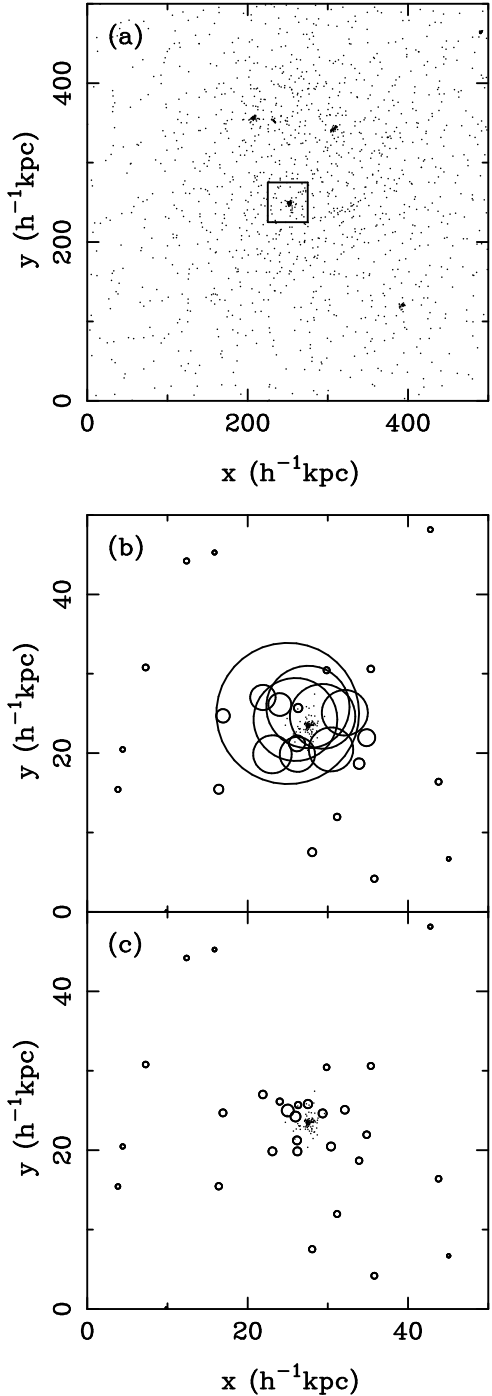

\centering
\PSbox{xemdots.ps angle=-90 voffset=580 hoffset=-265 vscale=100 hscale=100}
{3.5in}{7.3in} 
\caption
{
(a) A plot of particle positions in a $500 \hkpc$ cube centered on
the particle with the highest X-ray luminosity in the simulation.
(b) The central $ 50 \hkpc$ wide volume shown in the previous panel.
We plot each particle using a circle of
area proportional to its X-ray luminosity. 
(c) The same as panel (b), except that we have corrected the
density estimates by separating the hot and 
cold gas into two phases before calculating the X-ray luminosities.
\label{xemdots}
}
\end{figure}

In Figure \ref{xemdots} we show the positions and 
soft-band X-ray luminosities
of some particles before and after the density correction.
The panels are centered on the particle with the highest
uncorrected X-ray luminosity
($7 \times 10^{42} \ergs$), which resides in a group with
a circular velocity of $500 \kms$. Clumps of cold
particles are embedded in the intragroup gas, which, if included
in the density estimate of the hot particles, leads to the
artificially high X-ray emission seen in Figure \ref{xemdots}b.
After the correction, the X-ray luminosities of the 
particles near the galaxy are similar to those in the rest 
of the intragroup medium. The corrected soft-band luminosity of the group
is $1.8 \times 10^{42} \ergs$, a factor $12$ lower than the uncorrected 
value. 

With our fiducial parameters for excluding cold gas, we find
that the average volume emissivity of the whole simulation
volume is $1.55\times10^{-35} \ergvol$, and
$7.01\times10^{-36} \ergvol$, for the $0.5-2$ keV and  $2-10$ keV
bands, respectively. This is a factor of [13,7] smaller
than with the uncorrected density estimate, for the [soft,hard] band.
As a test that this correction is robust, we 
raise the cold density limit to $\rho >10^{5}$,
 and  find  $1.60\times10^{-35} \ergvol$,
$7.11\times10^{-36} \ergvol$ (for $0.5-2$, $2-10$ keV emissivities).
If we only change densities for hot gas if the hot gas ($T>10^5K$) density
was previously $>100$, we find (with cold $\rho$ limit of 1000)
 $1.55\times10^{-35} \ergvol$,
$7.07\times10^{-36} \ergvol$ (for $0.5-2$, $2-10$ keV).
 
These large changes in the thresholds for including cold
particles  and recalculating the densities of hot particles change the
corrected X-ray emission by $3 \%$ or less. This is a good indication
that our correction is sensible. This recalculation of
the density radically affects a small number of 
particles but has virtually no effect on the rest.
In the plot of temperature against density (Figure \ref{rhot}) we
used the corrected densities, as well as in  the results
in the main text of the paper.

\subsection{Isothermal sphere test}

As an additional test of our correction, we  set up a 
static toy model of a group.
This is to demonstrate that 
the spurious X-ray emission in the simulation can
be reproduced, and is simply due to the density kernel
estimation, something that can be tested without
hydrodynamics.

 We distribute 10,000 particles by randomly
sampling the density profile
\begin{equation}
\rho(r)\propto\frac{1}{r^{2}+r^{2}_{c}},
\end{equation}
to approximate an isothermal sphere with a core.
The parameter $r_{c}$ is set to $100 \hkpc$, and the temperature
of the particles to $10^{7}$ K. We add a top hat sphere
of radius $10 \hkpc$ containing 2500 particles to the
center of the group. These particles have a temperature of
$10^{4}$ K and represent a galaxy. We make sure that no hot
intragroup particles lie within the galaxy radius.

\begin{figure}[t]
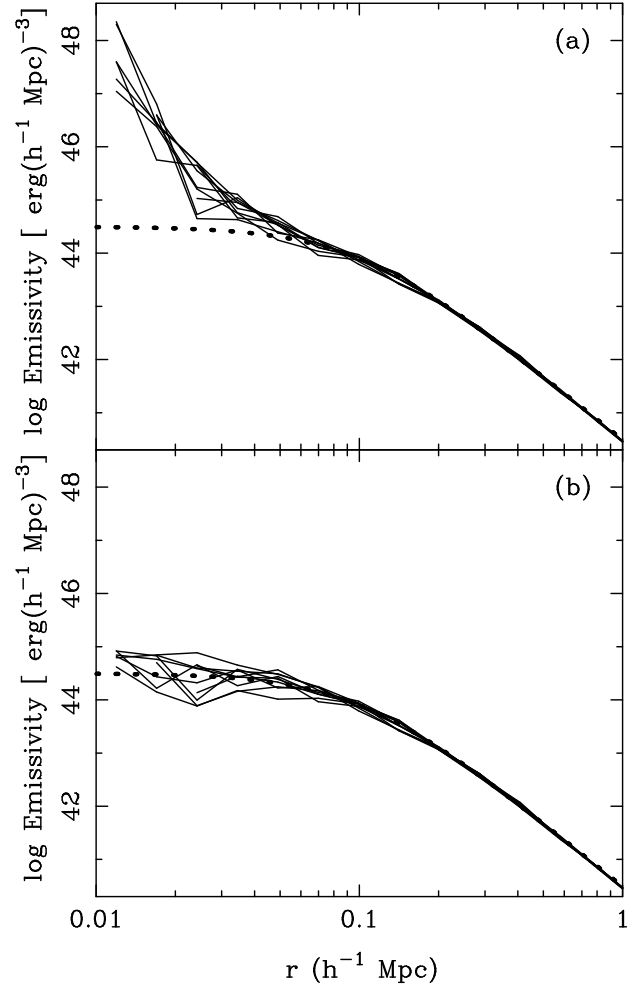

\centering
\PSbox{isothermal.ps angle=-90 voffset=430 hoffset=-175 vscale=75 hscale=75}
{3.5in}{5.3in} 
\caption
{
Solid lines: X-ray emission from isothermal spheres with an embedded
central galaxy: (a) Using the standard SPH density estimation, and
(b) after excluding cold particles from the density estimate
of hot particles.
The dashed line shows the correct analytic result. 
\label{isothermal}
}
\end{figure}

The X-ray emissivity profiles of 10 such toy groups
set up using different random seeds are shown in the 
top panel of Figure \ref{isothermal}, along with the analytic
profile. We can see that as we approach the central galaxy,
the emission rises rapidly above the correct result, due to the hot
particles near the center having their densities overestimated.
The total average luminosity of the groups is $1.1 \times 10^{43} \ergs$,
and there is a wide spread in individual values, with some having
luminosities a quarter of the mean, and some twice the mean.

If we do not include the galactic gas in the hot gas density estimate,
as would happen with our correction, the analytic profile is
well reproduced (Figure  \ref{isothermal}b). The corrected
mean luminosity is $2.9 \times 10^{42} \ergs$, with a maximum
$10\%$ variation between groups, due to shot noise.
We note that the difference between the uncorrected
and corrected luminosities (a factor $\sim 4$) is lower than 
that for the simulation group in Figure \ref{xemdots}. This is presumably
due to the fact that there is more than one galaxy in the simulation
group.

\subsection{Simulation resolution test}

With higher mass resolution, we would resolve the two distinct
gas phases better, so that the correction to the X-ray luminosities
that occurs when we separate them by hand should be smaller. 
Exploring the dependence of the correction on resolution is
therefore another way of checking its validity.

Although we cannot resimulate our entire volume at higher
resolution, a simulation in a much smaller 
box is available, one that was also used by D00.
 The cosmological model parameters are identical to our
 $50 \hmpc$ $\Lambda$CDM simulation, but the box
side-length is $11.11 \hmpc$, which gives a particle 
gas mass that is 8 times smaller, $1.1\times10^{8} \msun$.
The linear force resolution is also twice as high.

\begin{figure}[t]
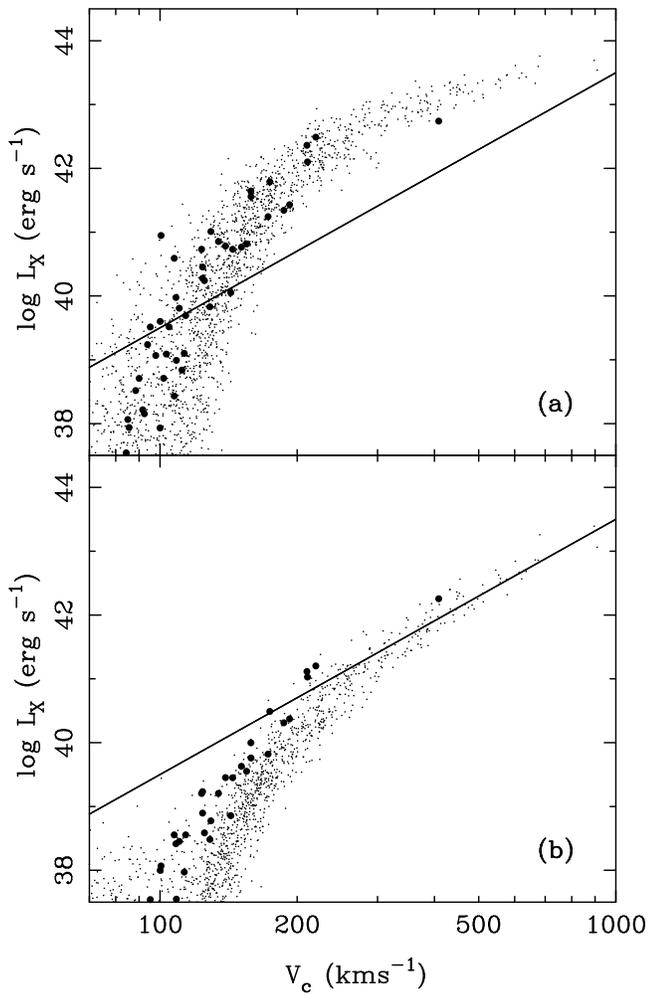

\centering
\PSbox{scatxemgrp.ps angle=-90 voffset=440 hoffset=-150 vscale=75 hscale=75}
{3.5in}{5.4in} 
\caption
{
The soft band X-ray luminosities of groups plotted
against their circular velocities, for two simulations
with different mass resolutions.
The dots represent groups in the large
simulation we have used in the main body of the paper,
and the large points groups taken from a much smaller
box with 8 times higher mass resolution (see text). 
 In panel (a) we show
the uncorrected luminosities, and (b) the 
corrected luminosities
that result when we separate the cold dense phase
from the hot gas.
The line $L_{X}\propto V_{c}^{4}$ is plotted with the same
(arbitrary) amplitude in each panel, to act as a reference line.
\label{scatxemgrp}
}
\end{figure}

To compare like with like, we do not consider the mean total
emission from the box (which in the smaller volume is affected
strongly by missing large-scale power), but compare galaxy groups of 
similar size taken from the two simulations. We have picked out
groups using a standard friends-of-friends algorithm
on the gas and dark matter particle distributions. The
circular velocity at the virial radius,
$V_{c}$, was estimated using a 
spherical overdensity criterion. In Figure \ref{scatxemgrp},
we plot the soft-band X-ray luminosities (we refer to the soft-band
for the rest of this subsection)
of these groups against $V_{c}$, for both the original
and corrected density estimates.

We concentrate on the largest group in the $11.11 \hmpc$ simulation
(with $V_{c}=410 \kms$),
because it is responsible for $74 \%$ of the (corrected)
 X-ray emission. Groups with this value of 
$V_{c}$ and higher account for $78 \%$ of the  X-ray emission in the
large box, so that tests carried out on such groups will
be representative of the total X-ray luminosity. 
{}From Figure \ref{scatxemgrp}a we can see that the high-resolution
group has a luminosity less than
 the other groups with similar $V_{c}$.
The  34 groups
in the large simulation with $V_{c}$ within $10\%$  of  $410 \kms$
have a mean uncorrected luminosity of $1.0\times10^{43} \ergs$, 
which is 6 times higher than that of the high-resolution group.
After the two phases have been separated
(Figure \ref{scatxemgrp}b), the mean luminosity 
of the low resolution groups
falls by a factor of 15, compared to 3 for the high resolution group.
 The latter is now on the high end, with luminosity some
$75 \%$ higher than the mean. 

With higher resolution, we find  that the correction
due to considering the two gas phases separately is smaller, as one
would expect. We would like more high resolution large groups, 
to make a good statistical comparison, but this is not possible.
 For the moment we can look 
at the small groups, with $V_{c} \sim 200 \kms$ and less. From
Figure \ref{scatxemgrp} we can see that they have much larger
correction factors than the large groups, and that the higher resolution
groups again have smaller corrections. The high-resolution groups
are also on average $3$ times more luminous than the others, although
as only $<5 \%$ of the total luminosity comes from  
groups below $V_{c} \sim 200 \kms$, this is not significant
for the total emission.

Another point of note is that the dispersion about the
mean luminosity for a given $V_{c}$ is noticeably smaller for
the corrected values. This is because the spurious
luminosities are affected by discreteness noise, being due to only a few
particles, as we have seen in the previous subsection.

The main aim of this resolution
test was to see whether the two-phase density correction has
a smaller effect at higher resolution. However, we would obviously
also like to study how the mean X-ray emissivity of the
simulation volume is affected
by resolution. We have found that the smallest groups,
which contribute only a few percent to the total, are affected by
resolution (as stated above,
a factor $3$ higher luminosity for a mass resolution $8$ times better).
Considering material more representative of the
bulk of the emission, the one large group at high resolution 
has $75 \%$ higher luminosity 
than the low resolution mean. It seems plausible that the 
difference may be even smaller for 
larger groups and clusters. Even if this is not the case,
it does not seem that the X-ray background intensity will be more
than mildly resolution
dependent. This is what might be expected if much of
the soft X-ray emitting material is in relatively 
diffuse gas, rather than being tightly bound to 
a hierarchy of small objects.

\subsection{Summary}

In this Appendix we have shown that because of the
way the SPH densities are estimated, we need to be careful with 
our treatment of
hot X-ray emitting gas 
 in close proximity to cold dense gas. 
With limited mass resolution, the densities
of a small number of hot particles are overestimated, something 
which has been realized in simulations of galaxy formation
(Pearce et al.\ 1999; Ritchie \& Thomas 2000),
and which
is especially important for the study of X-ray emission.
The simple solution to the problem is to separate
the gas phases before estimating the density and the X-ray
emissivity. We have done this, and carried out a number of
tests to check that the procedure is
both reasonable and robust. First, we have shown that the mean 
corrected X-ray emissivity varies by $<3\%$ when the 
density and temperature thresholds for separating 
the phases are varied by a factor of $100$. Second,
we have shown that the effect of two phases on the X-ray emission
can be reproduced and corrected in a controlled situation, an isothermal
sphere. Third, we have found that with higher mass
resolution, the correction factor is smaller, as we would expect.
Effects such as these are probably responsible for the steep central X-ray
profiles seen in simulations of galaxy clusters that have included cooling
(e.g. Katz \& White 1993; Tsai, Katz \& Bertschinger 1994; Lewis \etal 2000).
Using the corrected densities in the calculation of the X-ray emissivities
(see also Pearce \etal 2000)
may well bring these central regions into agreement with observations.

\end{document}